\documentclass[12pt]{article}

\newif\ifBlinded
\Blindedfalse

\newif\ifTitlePageOnly
\TitlePageOnlyfalse
\ifTitlePageOnly
  \Blindedfalse
\fi

\usepackage[margin=0.9in]{geometry}
\usepackage[T1]{fontenc}
\usepackage[utf8]{inputenc}
\usepackage{newpxtext}

\usepackage{mathtools}
\mathtoolsset{showonlyrefs}
\usepackage{newpxmath}
\usepackage{booktabs}
\usepackage{enumitem}
\usepackage{microtype}
\usepackage{bm}
\usepackage[super,sort&compress]{natbib}
\bibpunct{}{}{,}{s}{}{,}

\usepackage[dvipsnames]{xcolor}
\usepackage{tabularx}
\usepackage{graphicx}
\usepackage{caption}
\usepackage{subcaption}
\usepackage{float}
\usepackage{placeins}
\usepackage{pdflscape}

\usepackage{setspace}
\usepackage{titlesec}
\usepackage{titletoc}

\usepackage{afterpage}

\graphicspath{{Plots/}}

\floatstyle{ruled}
\newfloat{algorithm}{tbp}{loa}
\floatname{algorithm}{ALGORITHM}

\usepackage{hyperref}
\hypersetup{
  colorlinks = true,
  linkcolor = blue,
  citecolor = red,
  urlcolor = blue
}

\usepackage{siunitx}

\let\oldeqref\eqref
\renewcommand{\eqref}[1]{%
  \begingroup
  \hypersetup{linkcolor=black}%
  \oldeqref{#1}%
  \endgroup
}

\usepackage{doi}

\usepackage{orcidlink}

\usepackage{fancyhdr}
\usepackage{lastpage}


\newcommand{\mainsecref}[1]{\hyperref[#1]{Section~\ref*{#1}}}
\newcommand{\mainfigref}[1]{\hyperref[#1]{Figure~\ref*{#1}}}
\newcommand{\maintabref}[1]{\hyperref[#1]{Table~\ref*{#1}}}
\newcommand{\mainalgref}[1]{\hyperref[#1]{Algorithm~\ref*{#1}}}

\newcommand{\appref}[1]{\hyperref[#1]{Appendix~\ref*{#1}}}
\newcommand{\apptabref}[1]{\hyperref[#1]{Appendix Table~\ref*{#1}}}
\newcommand{\appfigref}[1]{\hyperref[#1]{Appendix Figure~\ref*{#1}}}
\newcommand{\appalgref}[1]{\hyperref[#1]{Algorithm~\ref*{#1}}}

\newcommand{\suppsecref}[1]{\hyperref[#1]{Section~\ref*{#1}}}
\newcommand{\supptabref}[1]{\hyperref[#1]{Table~\ref*{#1}}}
\newcommand{\suppfigref}[1]{\hyperref[#1]{Figure~\ref*{#1}}}
\newcommand{\suppalgref}[1]{\hyperref[#1]{Algorithm~\ref*{#1}}}
\newcommand{\suppeqref}[1]{\hyperref[#1]{Equation~\ref*{#1}}}

\newcommand{\supportingref}[1]{\hyperref[#1]{Supporting Information} (\hyperref[#1]{Section~\ref*{#1}})}

\newcommand{\E}{\mathbb{E}}
\newcommand{\Var}{\mathrm{Var}}

\newcommand{\Normal}{\mathcal{N}}

\newcommand{\R}{\mathbb{R}}

\newcommand{\ManuscriptTitle}{Reference-based multiple imputation for longitudinal bounded discrete outcomes}

\newcommand{\AuthorBlock}{
Divan Aristo Burger\,\orcidlink{0000-0001-8096-6371}$^{1,2,*}$
}

\newcommand{\AffiliationBlock}{
\begin{tabularx}{\textwidth}{@{}r@{\hspace{0.4em}}X@{}}
$^{1}$ & Cytel Inc., Waltham, MA, USA\\
$^{2}$ & Department of Mathematical Statistics and Actuarial Science, University of the Free State, Bloemfontein, South Africa
\end{tabularx}
}

\newcommand{\CorrespondingAuthor}{
\begin{tabularx}{\textwidth}{@{}r@{\hspace{0.4em}}X@{}}
$^{*}$ & Corresponding author: Divan A. Burger, Cytel Inc., 1050 Winter Street, Waltham, MA 02451, USA. Email: \href{mailto:divanaburger@gmail.com}{\texttt{divanaburger@gmail.com}}
\end{tabularx}
}

\newcommand{\MainAbstractText}{
Reference-based multiple imputation has been developed for binary and ordinal responses. We develop a parsimonious formulation for longitudinal bounded scores and counts that preserves their finite integer support and gives partial and complete reference borrowing an explicit distributional interpretation. Discrete Kumaraswamy margins provide closed-form cumulative distribution and quantile functions, simplifying fitting and imputation under a Gaussian copula. Borrowing is defined through a regression model indexed by a smooth median anchor, allowing continuation under the assigned treatment, partial borrowing, or jump to reference after a specified trigger. With shapes shared across arms, complete borrowing replaces the active marginal distribution with the reference distribution at the same visit and covariate values, while conditional imputations remain informed by retained outcomes. Bayesian estimation uses latent variable augmentation, and all sensitivity rules share one fitted posterior. Completed datasets are analyzed using a beta-binomial random intercept model, with treatment comparisons expressed as standardized mean score differences. Simulations assess parameter recovery and imputation followed by analysis. An application to depression scores illustrates attenuation of the treatment benefit at the final visit as borrowing increases, with confidence intervals including zero under all rules.
}

\newcommand{\KeywordBlock}{
\begin{tabularx}{\textwidth}{@{}l@{\hspace{0.5em}}X@{}}
\textbf{KEYWORDS:} & Bounded discrete outcomes; Gaussian copula; Kumaraswamy distribution; Missing data; Reference-based multiple imputation; Sensitivity analysis
\end{tabularx}
}

\newcommand{\MainTitlePageContent}{
\noindent\textbf{ABSTRACT}

\vspace{0.25cm}

\noindent\ignorespaces\MainAbstractText

\vspace{0.5cm}

\KeywordBlock
}

\newcommand{\MakeManuscriptTitlePage}[2]{
\begin{titlepage}
\thispagestyle{empty}

{\Large\bfseries #1\par}

\vspace{0.5cm}

\rule{\textwidth}{0.4pt}

\ifBlinded
\vspace{0.5cm}
\else
\vspace{0.6cm}

{\large \AuthorBlock\par}

\vspace{0.55cm}

\begin{minipage}{\textwidth}
{\small \AffiliationBlock\par}
\end{minipage}

\vspace{0.45cm}

\begin{minipage}{\textwidth}
{\normalsize \CorrespondingAuthor\par}
\end{minipage}

\vspace{0.5cm}
\fi

\begin{minipage}{\textwidth}
#2
\end{minipage}

\end{titlepage}
\newpage
}

\newcommand{\EthicsStatement}{%
This methodological study used simulated data and a secondary analysis of the publicly available \texttt{BtheB} dataset distributed in the R package \texttt{HSAUR2}. No participants were recruited, no new participant data were collected, and no directly identifying information was accessed. No additional ethics approval was obtained for this secondary analysis.
}

\newcommand{\ParticipationStatement}{%
No new patient consent was sought for this secondary analysis of publicly available data. The study involved no participant contact, and no identifiable participant information is reported.
}

\newcommand{\PublicationConsentStatement}{%
Not applicable. No identifiable individual information is reported.
}

\newcommand{\ConflictStatement}{%
The author is employed by Cytel Inc. and declares no other conflicts of interest relevant to this work.
}

\newcommand{\FundingStatement}{%
No funding was received for this research.
}

\newcommand{\TitlePageDeclarations}{%
\section*{Statements and declarations}
\subsection*{Ethical considerations}
\EthicsStatement
\subsection*{Consent to participate}
\ParticipationStatement
\subsection*{Consent for publication}
\PublicationConsentStatement
\subsection*{Declaration of conflicting interest}
\ConflictStatement
\subsection*{Funding}
\FundingStatement
\subsection*{Data availability}
The application data are available as \texttt{BtheB} in the R package \texttt{HSAUR2}. R code for the application and simulation studies is available at \url{https://github.com/DABURGER1/RBMI-Discrete-Bounded}.
}

\newcommand{\MakeSubmissionTitlePage}{%
\pagestyle{empty}
{\Large\bfseries \ManuscriptTitle\par}
\vspace{0.5cm}
\rule{\textwidth}{0.4pt}
\vspace{0.6cm}
{\large \AuthorBlock\par}
\vspace{0.55cm}
{\small \AffiliationBlock\par}
\vspace{0.45cm}
\CorrespondingAuthor\par
\TitlePageDeclarations
}

\newcommand{\SupplementaryContents}{%
{\large\bfseries Contents\par}
\medskip
\printcontents[supplement]{}{1}{\setcounter{tocdepth}{2}}%
}

\makeatletter
\newcommand{\StartSupplementaryMaterial}{%
\clearpage
\startcontents[supplement]
\setcounter{section}{0}
\setcounter{subsection}{0}
\setcounter{subsubsection}{0}
\setcounter{figure}{0}
\setcounter{table}{0}
\setcounter{algorithm}{0}
\setcounter{equation}{0}
\renewcommand{\thesection}{S\arabic{section}}
\renewcommand{\thefigure}{S\arabic{figure}}
\renewcommand{\thetable}{S\arabic{table}}
\renewcommand{\thealgorithm}{S\arabic{algorithm}}
\renewcommand{\theequation}{S\arabic{equation}}
\let\Hy@chapapp\Hy@chapterstring
\renewcommand{\theHsection}{supplement.\arabic{section}}
\renewcommand{\theHsubsection}{\theHsection.\arabic{subsection}}
\renewcommand{\theHsubsubsection}{\theHsubsection.\arabic{subsubsection}}
\renewcommand{\theHfigure}{supplement.\arabic{figure}}
\renewcommand{\theHtable}{supplement.\arabic{table}}
\def\theHalgorithm{supplement.\arabic{algorithm}}
\renewcommand{\theHequation}{supplement.\arabic{equation}}
\titleformat{\section}
  {\normalfont\Large\bfseries}
  {\thesection}
  {1em}
  {}
\captionsetup[figure]{name = Figure}
\captionsetup[table]{name = Table}
\floatname{algorithm}{Algorithm}
\captionsetup*[algorithm]{name = Algorithm}
\hypersetup{pageanchor=false}
\MakeManuscriptTitlePage
  {\pdfbookmark[0]{Supplementary Material}{supplementary-material}%
   Supplementary Material\par\vspace{0.35cm}{\large\ManuscriptTitle}}
  {\SupplementaryContents}
\pagenumbering{arabic}
\renewcommand{\thepage}{S\arabic{page}}
\hypersetup{pageanchor=true}
}
\makeatother

\newcommand{\expit}{\operatorname{expit}}
\newcommand{\logit}{\operatorname{logit}}
\newcommand{\ind}{\mathbb{I}}

\hypersetup{pdftitle={\ManuscriptTitle}}
\ifBlinded
  \hypersetup{pdfauthor={}}
\else
  \hypersetup{pdfauthor={Divan A. Burger}}
\fi

\begin{document}
\ifTitlePageOnly
\hypersetup{pageanchor=false}
\MakeSubmissionTitlePage
\else
\hypersetup{pageanchor=false}
\MakeManuscriptTitlePage{\ManuscriptTitle}{\MainTitlePageContent}
\hypersetup{pageanchor=true}

\section{Introduction}
\label{sec:introduction}

Bounded discrete outcomes arise in clinical trials as symptom or functioning scores and as counts of successes among a fixed number of opportunities. Such outcomes may be skewed, show substantial variation between participants, and accumulate near one endpoint. Missing follow-up measurements introduce a further difficulty: conclusions may depend on unverifiable assumptions about the unobserved outcomes, particularly after treatment discontinuation or another clinically relevant event. An imputation model that respects the outcome's finite support and uses the participant's observed trajectory provides a natural basis for sensitivity analysis in this setting.

Reference-based multiple imputation provides a framework for exploring departures from missing at random (MAR) by specifying assumptions about unobserved outcomes using information from a reference treatment group.\cite{carpenter2013,cro2020} For missing outcomes in the active arm after a specified trigger, continuation under the assigned treatment retains the fitted distribution for that arm, whereas jump to reference (J2R) borrows features of the distribution in the reference arm. Intermediate assumptions can retain a proportion of the fitted treatment contrast on a specified model scale. Controlled imputation has been developed for several outcome types, including recurrent events,\cite{keene2014} binary responses, and ordinal responses.\cite{cro2025binary,tang2018ordinal} The present focus is on bounded scores and counts. Although such outcomes can be treated as ordinal, a parametric margin with a fixed number of parameters provides a parsimonious alternative to specifying a separate threshold for each boundary between categories.

Reference-based imputation for repeated counts has been developed using negative binomial margins and a Gaussian copula.\cite{burger2026nb} The construction separates assumptions about the marginal outcomes after the trigger from the dependence model used to condition on retained observations. Negative binomial margins, however, assign positive probability above any finite upper bound. Replacing out-of-range imputations with the maximum would change the specified distribution and create artificial mass at the boundary. Extending this framework to bounded outcomes therefore motivates a marginal distribution that respects both endpoints of the outcome's support.

The discrete Kumaraswamy (DK) distribution\cite{barbiero2026dk} is a suitable choice of bounded margin, with an elementary closed-form cumulative distribution function (CDF) and an explicit quantile function. These features are particularly convenient in a copula model, where marginal CDFs are evaluated repeatedly during fitting and conditional imputation maps latent probabilities to outcome quantiles. A beta-binomial (BB) margin is a natural alternative, but its CDF typically requires finite summation or special-function evaluation, and its quantiles generally require a search over the support. The DK margin provides direct expressions for both operations, with its quantile obtained using a ceiling operation, simplifying the marginal calculations within the copula algorithm.

We formulate reference-based multiple imputation for bounded discrete outcomes by extending the count framework\cite{burger2026nb} with DK margins indexed by a smooth median anchor and a shape parameter. The formulation defines partial and complete reference borrowing through the treatment contrast on the transformed anchor scale. With shape parameters shared across arms, complete borrowing replaces the marginal distribution in the active arm after the trigger with that in the reference arm at the same visit and covariate values, while conditional imputations remain informed by the participant's retained outcomes. The Gaussian copula and augmentation with latent variables follow existing work. We illustrate the procedure with depression scores from the public BtheB dataset,\cite{hsaur2} analyzing the completed datasets using a BB working model with a random intercept. Treatment comparisons use mean scores standardized to a common covariate distribution after integration over the random intercept distribution. Thus, the analysis targets the mean, while the median anchor parameterizes the marginal distributions used for imputation.

The remainder of the paper is organized as follows. \autoref{sec:methods} presents the model, reference-based imputation rules, posterior computation, and analysis of completed datasets. \autoref{sec:application} illustrates the approach using the BtheB data. \autoref{sec:simulation} examines parameter recovery and imputation followed by analysis, and \autoref{sec:discussion} discusses the implications and limitations of the approach.

\section{Methods}
\label{sec:methods}

\subsection{Data structure and the imputation trigger}
\label{sec:missingness}

Let $Y_{ij} \in \left\{0, \ldots, K\right\}$ be the outcome for participant $i = 1, \ldots, n$ at scheduled time $t_j$, where $j \in \mathcal V = \left\{1, \ldots, J\right\}$, $t_1 < \cdots < t_J$, and $K \geq 2$ is a known integer. Let $A_i \in \left\{0, 1\right\}$ indicate reference or active treatment, respectively. Let $\bm x_i = \left(x_{i1}, \ldots, x_{ip}\right)^\top \in \R^p$ contain $p \geq 1$ fully observed pretreatment covariates, with their coding and scaling specified before fitting. The vector excludes the intercept and treatment indicator, which enter the model separately. Participants are independent conditional on the model parameters, treatment, and these covariates. Conditioning on treatment and $\bm x_i$ is suppressed in subsequent notation.

Let $\mathcal R_i \subseteq \mathcal V$ contain the visits with recorded outcomes, and let $D_i \in \left\{1, \ldots, J + 1\right\}$ be the first visit to which a post-trigger assumption applies. The value $D_i = J + 1$ places the trigger beyond scheduled follow-up. Following the count framework,\cite{burger2026nb} define
\begin{equation}
\mathcal O_i = \left\{j \in \mathcal R_i : j < D_i\right\},
\qquad
\mathcal M_i = \mathcal V \setminus \mathcal O_i.
\label{eq:retained-sets}
\end{equation}
Write $y_{ij}$ for a recorded realization of $Y_{ij}$. The outcomes at visits in $\mathcal O_i$ are retained for model fitting and conditioning; visits in $\mathcal M_i$ require completion (imputation). Missing outcomes before $D_i$ are imputed under the model for the assigned treatment, conditional on all retained outcomes, including those at later visits. All visits at or after $D_i$ use the selected post-trigger rule. Thus, if a clinical event defines the trigger, recorded outcomes on or after that event are set aside under this convention. That choice must accord with the scientific target. The trigger and retained set are specified before fitting.

\subsection{Discrete Kumaraswamy margins indexed by a median anchor}
\label{sec:margins}

For positive shapes $a$ and $b$, the shifted DK distribution\cite{barbiero2026dk} has CDF
\begin{equation}
F_K\left(y \middle| a, b\right)
= 1 - \left\{1 - \left(\frac{y + 1}{K + 1}\right)^a\right\}^{b},
\qquad y = 0, \ldots, K.
\label{eq:cdf}
\end{equation}
Set $F_K\left(y \middle| a, b\right) = 0$ for $y < 0$ and $F_K\left(y \middle| a, b\right) = 1$ for $y \geq K$; between successive integers, the CDF is constant. The probability mass function (PMF) is
\begin{equation}
p_K\left(y \middle| a, b\right)
= F_K\left(y \middle| a, b\right) - F_K\left(y - 1 \middle| a, b\right),
\qquad y = 0, \ldots, K.
\label{eq:pmf}
\end{equation}
These probabilities are positive at every support point, including both endpoints, and sum to one.

For participant $i$ at visit $j$, specify the linear predictor
\begin{equation}
\eta_{ij} = \alpha_j + \beta_j A_i + \bm x_i^\top\bm\gamma_j,
\label{eq:regression}
\end{equation}
where $\alpha_j$ and $\beta_j$ are the visit-specific intercept and treatment coefficient. The vector $\bm\gamma_j = \left(\gamma_{j1}, \ldots, \gamma_{jp}\right)^\top$ contains the visit-specific covariate coefficients, which are shared across treatment arms. Using the quantile and shape parameterization of the continuous Kumaraswamy family\cite{bayes2017} at probability level $1/2$, define
\begin{equation}
\begin{aligned}
r_{ij} &= \expit\left(\eta_{ij}\right), &
\xi_{ij} &= \left(K + 1\right)r_{ij} - 1,\\
a_{ij} &= -\frac{\phi_j}{\log r_{ij}}, &
b_j &= \frac{\log\left(1/2\right)}{\log\left\{1 - \exp\left(-\phi_j\right)\right\}},
\qquad \phi_j > 0.
\end{aligned}
\label{eq:parameterization}
\end{equation}
Here, $\expit\left(x\right) = \left\{1 + \exp\left(-x\right)\right\}^{-1}$ and its inverse is $\logit\left(r\right) = \log\left\{r/\left(1 - r\right)\right\}$. The free shape parameter $\phi_j$ varies by visit and is shared across treatment arms. Write $F_{ij}\left(y\right) = F_K\left(y \middle| a_{ij}, b_j\right)$ and $p_{ij}\left(y\right) = p_K\left(y \middle| a_{ij}, b_j\right)$ for the resulting marginal CDF and PMF.

To interpret $\xi_{ij}$, let $H_K\left(x \middle| a, b\right)$ be the continuous CDF obtained by evaluating the analytic expression in \eqref{eq:cdf} for real $-1 \leq x \leq K$, with values zero and one outside this interval. Using a continuous interpolation to define a smooth quantile target for a discrete outcome is consistent with a model-aware approach.\cite{padellini2018model} Because $r_{ij}^{a_{ij}} = \exp\left(-\phi_j\right)$, the parameterization gives $H_K\left(\xi_{ij} \middle| a_{ij}, b_j\right) = 1/2$. Thus, $\xi_{ij} \in \left(-1, K\right)$ is a smooth median anchor and $r_{ij} = \left(\xi_{ij} + 1\right)/\left(K + 1\right)$ is its normalized value. For $0 < u < 1$, let $Q_{ij}\left(u\right)$ denote the lower quantile, defined as the smallest $y \in \left\{0, \ldots, K\right\}$ satisfying $F_{ij}\left(y\right) \geq u$. Since $H_K$ agrees with $F_K$ at integer support points, the discrete lower median is
\begin{equation}
Q_{ij}\left(1/2\right) = \left\lceil\xi_{ij}\right\rceil,
\label{eq:median}
\end{equation}
where $\left\lceil x\right\rceil$ is the smallest integer greater than or equal to $x$. The integer median can therefore remain unchanged as the anchor varies; anchors in $\left(-1, 0\right]$ give an integer median of zero. At a fixed anchor, $\phi_j$ controls the remaining distributional shape and is not itself a variance.

The general quantile has the explicit expression
\begin{equation}
Q_{ij}\left(u\right)
= \left\lceil\left(K + 1\right)
\left\{1 - \left(1 - u\right)^{1/b_j}\right\}^{1/a_{ij}} - 1\right\rceil,
\qquad 0 < u < 1.
\label{eq:inverse}
\end{equation}
It returns an integer in $\left\{0, \ldots, K\right\}$ directly. The derivation is given in \appref{app:ordering}. The regression describes each visit's marginal distribution conditional on treatment and the covariates, before conditioning on retained follow-up outcomes. Although indexed by a median anchor, the model defines the entire distribution needed for imputation. At fixed parameter values, its marginal mean is
\begin{equation}
\E\left(Y_{ij}\right)
= \sum_{y=0}^{K} y p_{ij}\left(y\right)
= \sum_{y=0}^{K-1}\left\{1 - F_{ij}\left(y\right)\right\},
\label{eq:dk-mean}
\end{equation}
with conditioning on treatment and the covariates suppressed as before. Averaging these means over a common covariate distribution gives standardized means under the marginal model for the assigned treatment. The analysis of completed datasets in \mainsecref{sec:analysis} estimates means from its own fitted model.

\subsection{Gaussian copula and treatment of discreteness}
\label{sec:copula}

Let $\rho \in \left(0, 1\right)$ govern longitudinal dependence, and let $\bm\theta$ collect the $J\left(p + 3\right) + 1$ parameters comprising $\alpha_j$, $\beta_j$, the $p$ components of $\bm\gamma_j$, and $\log\phi_j$ at each visit, together with $\logit\rho$. The joint model under the assigned treatment is
\begin{equation}
\begin{aligned}
\left.\bm Z_i \middle| \bm\theta\right.
&\sim \Normal_J\left(\bm 0, \bm R\right),
& Y_{ij} &= Q_{ij}\left\{\Phi\left(Z_{ij}\right)\right\},\\
R_{jk} &= \rho^{\left|t_j - t_k\right|},
& j, k &\in \mathcal V,
\end{aligned}
\label{eq:copula}
\end{equation}
where $\bm Z_i = \left(Z_{i1}, \ldots, Z_{iJ}\right)^\top$, $\Phi$ is the standard normal CDF, and $\Normal_d\left(\bm m, \bm V\right)$ denotes the $d$-dimensional normal distribution with mean $\bm m$ and covariance $\bm V$. The parameter $\rho$ is the latent correlation between measurements separated by one unit of time. This continuous-time AR(1) structure accommodates unequal visit spacing and is common across arms. The entries of $\bm R$ describe correlations between the latent Gaussian variables. Correlations between observed scores also depend on the marginal distributions at each visit and can therefore differ across treatment arms. For index sets $S, T \subseteq \mathcal V$, $\bm R_{ST}$ denotes the submatrix with rows in $S$ and columns in $T$, ordered by visit.

A retained score $y_{ij}$ is equivalent to the latent interval constraint
\begin{equation}
\Phi^{-1}\left\{F_{ij}\left(y_{ij} - 1\right)\right\}
< Z_{ij} \leq
\Phi^{-1}\left\{F_{ij}\left(y_{ij}\right)\right\},
\qquad j \in \mathcal O_i.
\label{eq:latent-interval}
\end{equation}
The lower endpoint is $-\infty$ when $y_{ij} = 0$, and the upper endpoint is $+\infty$ when $y_{ij} = K$. Let $d_i = \left|\mathcal O_i\right|$. For $d_i > 0$, let $L_i\left(\bm\theta\right)$ be the probability of all these constraints under $\Normal_{d_i}\left(\bm 0, \bm R_{\mathcal O_i\mathcal O_i}\right)$; for $d_i = 0$, set $L_i\left(\bm\theta\right) = 1$. The fitting likelihood is $L\left(\bm\theta\right) = \prod_{i=1}^{n} L_i\left(\bm\theta\right)$. The likelihood uses only retained outcomes. For fitting, the retention mechanism is treated as ignorable under a working MAR assumption, with its parameters taken to be distinct from, and a priori independent of, $\bm\theta$.

To sample the posterior without repeatedly evaluating these Gaussian rectangle probabilities, use latent variable augmentation,\cite{smith2012} as in the count framework.\cite{burger2026nb} Introduce $v_{ij} \in \left(0, 1\right)$ for each retained score and set
\begin{equation}
u_{ij} = F_{ij}\left(y_{ij} - 1\right) + v_{ij}p_{ij}\left(y_{ij}\right),
\qquad z_{ij} = \Phi^{-1}\left(u_{ij}\right).
\label{eq:pit}
\end{equation}
Let $\bm v_i$ and $\bm u_i$ collect these quantities over $j \in \mathcal O_i$ in visit order, and let $c_{\bm R_{\mathcal O_i\mathcal O_i}}$ be the Gaussian copula density with correlation matrix $\bm R_{\mathcal O_i\mathcal O_i}$. The augmented likelihood contribution is
\begin{equation}
L_i^{\mathrm{aug}}\left(\bm\theta, \bm v_i\right)
= c_{\bm R_{\mathcal O_i\mathcal O_i}}\left(\bm u_i\right)
\prod_{j \in \mathcal O_i} p_{ij}\left(y_{ij}\right),
\label{eq:augmentation}
\end{equation}
and its integral over $\left(0, 1\right)^{d_i}$ is exactly $L_i\left(\bm\theta\right)$. For $d_i = 1$, the copula density is one; for $d_i = 0$, the likelihood contribution is one, and there are no auxiliary variables. The latent construction defines a valid joint model, although a joint distribution with discrete margins does not generally determine a unique copula.

Writing $\mathcal Y_{\mathrm{ret}}$ for all retained scores, $\bm v$ for all auxiliary variables, and $p\left(\bm\theta\right)$ for the prior density specified in \mainsecref{sec:computation}, the joint posterior is
\begin{equation}
p\left(\bm\theta, \bm v \middle| \mathcal Y_{\mathrm{ret}}\right)
\propto p\left(\bm\theta\right)
\prod_{i=1}^{n} L_i^{\mathrm{aug}}\left(\bm\theta, \bm v_i\right).
\label{eq:posterior}
\end{equation}
The auxiliary variables have a uniform base measure, but their posterior distribution is weighted by the copula. Sampling them with the parameters integrates over the latent positions compatible with each score; fixing them at the midpoint or arbitrary randomized values would not perform this integration.

\subsection{Reference-based rules and conditional imputation}
\label{sec:rules}

For a prespecified borrowing weight $\omega \in \left[0, 1\right]$, define
\begin{equation}
\eta_{ij}^{\left(\omega\right)}
= \alpha_j + \bm x_i^\top\bm\gamma_j
+ \left\{1 - \omega\ind\left(j \geq D_i\right)\right\}\beta_j A_i,
\qquad j \in \mathcal V,
\label{eq:rule}
\end{equation}
where $\ind\left(E\right)$ is one when condition $E$ holds and zero otherwise. The weight is the proportion of the treatment coefficient removed after the trigger. Define $F_{ij}^{\left(\omega\right)}$ and $Q_{ij}^{\left(\omega\right)}$ by replacing $\eta_{ij}$ with $\eta_{ij}^{\left(\omega\right)}$ in \mainsecref{sec:margins}, while retaining $\phi_j$. The corresponding median anchor is
\begin{equation}
\xi_{ij}^{\left(\omega\right)}
= \left(K + 1\right)\expit\left(\eta_{ij}^{\left(\omega\right)}\right) - 1.
\label{eq:rule-anchor}
\end{equation}
These margins are joined by the same Gaussian copula as the model under the assigned treatment. The margins in the reference arm and all margins before the trigger remain unchanged.

Continuation under the assigned treatment uses $\omega = 0$, giving the conditional distribution of outcomes requiring completion under the fitted joint model for that treatment. This is the MAR benchmark: given treatment, the covariates, and retained outcomes, no additional dependence of these outcomes on the retention pattern is introduced. Halfway borrowing uses $\omega = 1/2$, and J2R uses $\omega = 1$. Halfway borrowing takes the midpoint between the active and reference predictors on the logit scale of the normalized median anchor; this does not necessarily halve a treatment difference on the original score scale.

Under J2R, the median anchor in the active arm after the trigger is replaced by that in the reference arm at the same visit and covariate values. Specifically, write $\xi_{ij}^{\mathrm{ref}}$ and $F_{ij}^{\mathrm{ref}}$ for the anchor and CDF obtained by setting $A_i = 0$ in the fitted model. For $A_i = 1$ and $j \geq D_i$,
\begin{equation}
\begin{aligned}
\xi_{ij}^{\left(1\right)}
&= \left(K + 1\right)\expit\left(\alpha_j + \bm x_i^\top\bm\gamma_j\right) - 1
= \xi_{ij}^{\mathrm{ref}},\\
F_{ij}^{\left(1\right)}\left(y\right)
&= F_{ij}^{\mathrm{ref}}\left(y\right),
\qquad y = 0, \ldots, K.
\end{aligned}
\label{eq:j2r-margin}
\end{equation}
The CDF equality follows because $\phi_j$ is shared across arms. Thus, J2R replaces the entire marginal distribution in the working joint model used for imputation. These equalities concern margins before conditioning on retained follow-up outcomes; the conditional distribution of an imputed score still depends on the active participant's history. Retaining $\phi_j$ and $\bm R$ across rules does not hold variances or covariances on the original score scale fixed.

Because the margins and copula on $\mathcal O_i$ are unchanged, the likelihood of the retained outcomes is identical across rules, so all rules use the same posterior in \eqref{eq:posterior}. Fit that posterior once. For each joint draw of $\bm\theta$ and the retained latent values $\bm Z_{i,\mathcal O_i}$ obtained from \eqref{eq:pit}, draw the components requiring completion from
\begin{equation}
\left.\bm Z_{i,\mathcal M_i} \middle| \bm Z_{i,\mathcal O_i}, \bm\theta\right.
\sim \Normal_{\left|\mathcal M_i\right|}\left(\bm m_i, \bm V_i\right),
\label{eq:conditional}
\end{equation}
where
\begin{equation}
\begin{aligned}
\bm m_i &= \bm R_{\mathcal M_i\mathcal O_i}
\bm R_{\mathcal O_i\mathcal O_i}^{-1}\bm Z_{i,\mathcal O_i},\\
\bm V_i &= \bm R_{\mathcal M_i\mathcal M_i}
- \bm R_{\mathcal M_i\mathcal O_i}\bm R_{\mathcal O_i\mathcal O_i}^{-1}
\bm R_{\mathcal O_i\mathcal M_i}.
\end{aligned}
\end{equation}
All components in $\mathcal M_i$ are drawn jointly, including intermittent gaps, and participants are sampled independently conditional on the posterior draw. If $\mathcal O_i$ is empty, use $\bm m_i = \bm 0$ and $\bm V_i = \bm R_{\mathcal M_i\mathcal M_i}$; if $\mathcal M_i$ is empty, no imputation is needed. The retained latent values are still obtained using the margins for the assigned treatment.

For each rule, construct the completed trajectory as
\begin{equation}
Y_{ij}^{\left(\omega\right)} =
\begin{cases}
y_{ij}, & j \in \mathcal O_i,\\
Q_{ij}^{\left(\omega\right)}\left\{\Phi\left(Z_{ij}\right)\right\},
& j \in \mathcal M_i.
\end{cases}
\label{eq:completion}
\end{equation}
This procedure propagates uncertainty in the parameters, retained latent positions, and outcomes requiring completion. The same latent draws for components requiring completion are reused across rules to reduce Monte Carlo noise in comparisons. This makes imputations in the reference arm and those before the trigger identical across rules. For fixed $\phi_j$, DK quantiles are nondecreasing in $\eta$. Hence, for an active participant after the trigger, a draw with $\beta_j < 0$ gives $Y_{ij}^{\left(0\right)} \leq Y_{ij}^{\left(1/2\right)} \leq Y_{ij}^{\left(1\right)}$; the inequalities reverse when $\beta_j > 0$. This ordering concerns paired imputations and does not necessarily carry over to pooled treatment effect estimates; see \appref{app:ordering}.

\mainalgref{alg:impute} summarizes the full procedure, including the analysis and pooling of completed datasets described in \mainsecref{sec:analysis}.

\begin{algorithm}[!htbp]
\caption{Reference-based imputation and analysis for bounded discrete outcomes}
\label{alg:impute}
\small
\setlength{\parskip}{0pt}
\vspace{4pt}
\noindent\textbf{Inputs:} Recorded outcomes, treatment and pretreatment covariates, visit times,
support bound $K$, triggers $D_i$, retained sets $\mathcal O_i$, and selected
borrowing weights $\omega$.
\begin{enumerate}[
  label = \arabic*.,
  leftmargin = 1.7em,
  labelsep = 0.5em,
  itemsep = 5pt,
  parsep = 0pt,
  topsep = 6pt
]
\item \textbf{Fit once.} Sample the joint posterior in \eqref{eq:posterior}
using the DK Gaussian copula model under the assigned treatment, retaining all post-warmup draws for posterior summaries and subsequent imputation.

\item \textbf{Complete datasets.} For each selected post-warmup draw of
$\left(\bm\theta, \bm v\right)$:
\begin{enumerate}[
  label = (\alph*),
  leftmargin = 2em,
  labelsep = 0.5em,
  itemsep = 3pt,
  parsep = 0pt,
  topsep = 3pt
]
\item Recover each participant's retained latent values
$\bm Z_{i,\mathcal O_i}$ using \eqref{eq:pit}.
\item For each participant, jointly draw $\bm Z_{i,\mathcal M_i}$
from the conditional Gaussian distribution in \eqref{eq:conditional}.
\item For each selected borrowing weight $\omega$, construct a completed
dataset using \eqref{eq:rule} and \eqref{eq:completion}. Reuse the same latent
draws across rules and preserve all retained scores.
\end{enumerate}

\item \textbf{Analyze and pool.} Select $M \geq 2$ completed datasets per rule
using a prespecified selection of draw indices that does not depend on the
imputed outcomes or analysis results. Use the same indices across rules.
Fit the prespecified analysis to each selected dataset, calculate the
standardized means, treatment differences, and their variances, and pool
separately within each rule as described in \mainsecref{sec:analysis}.
\end{enumerate}
\textbf{Output:} One completed dataset per selected joint draw and rule, with
treatment estimates and intervals pooled separately for each rule.
\par\vspace{4pt}
\end{algorithm}

\subsection{Posterior computation and checking}
\label{sec:computation}

Independent priors are assigned to the regression coefficients, shape parameters, and correlation:
\begin{equation}
\begin{aligned}
\alpha_j &\sim \Normal\left(\mu_\alpha, \sigma_\alpha^2\right), &
\beta_j &\sim \Normal\left(\mu_\beta, \sigma_\beta^2\right),\\
\gamma_{jk} &\sim \Normal\left(\mu_{\gamma k}, \sigma_{\gamma k}^2\right), &
\log\phi_j &\sim \Normal\left(\mu_\phi, \sigma_\phi^2\right),\\
\rho &\sim \operatorname{Beta}\left(a_\rho, b_\rho\right).
\end{aligned}
\label{eq:priors}
\end{equation}
Here, $j = 1, \ldots, J$, $k = 1, \ldots, p$, and the second argument of $\Normal$ is the variance. The prior means $\mu_\alpha$, $\mu_\beta$, $\mu_{\gamma k}$, and $\mu_\phi$ are real-valued fixed hyperparameters. The corresponding standard deviations (SDs) $\sigma_\alpha$, $\sigma_\beta$, $\sigma_{\gamma k}$, and $\sigma_\phi$, and the beta shape hyperparameters $a_\rho$ and $b_\rho$, are positive. These hyperparameters are specified before fitting and are shared across visits where applicable. The hyperparameters for $\gamma_{jk}$ may differ across covariates to reflect their coding and scales. The shape prior is specified directly on $\log\phi_j$, whereas the correlation prior is specified on $\rho$. The correlation prior refers to a separation of one unit on the time scale used for $t_j$.

The parameter vector $\bm\theta$ uses $\zeta = \logit\rho$ as its correlation coordinate. Its prior contribution therefore includes the transformation Jacobian:
\begin{equation}
p_\zeta\left(\zeta\right)
= \frac{\rho^{a_\rho - 1}\left(1 - \rho\right)^{b_\rho - 1}}
{\mathrm{B}\left(a_\rho, b_\rho\right)}
\rho\left(1 - \rho\right)
= \frac{\rho^{a_\rho}\left(1 - \rho\right)^{b_\rho}}
{\mathrm{B}\left(a_\rho, b_\rho\right)},
\qquad \rho = \expit\left(\zeta\right),
\label{eq:rho-prior-coordinate}
\end{equation}
where $\mathrm{B}\left(a_\rho, b_\rho\right)$ is the beta function evaluated at the two prior shape hyperparameters. This contribution is included in $p\left(\bm\theta\right)$ in \eqref{eq:posterior}. No additional Jacobian is needed for the shape prior because both that prior and the sampling coordinate use $\log\phi_j$.

A custom Metropolis-within-Gibbs sampler uses Gaussian random-walk proposals to update $\alpha_j$, $\beta_j$, $\bm\gamma_j$, and $\log\phi_j$ jointly for each visit and $\logit\rho$ in a separate block. For the augmentation, proposals are made on $w_{ij} = \logit\left(v_{ij}\right)$, jointly over each participant's retained visits. Consequently, the target density on these coordinates includes the Jacobian
\begin{equation}
\prod_{i=1}^{n}\prod_{j \in \mathcal O_i}
v_{ij}\left(1 - v_{ij}\right)
\end{equation}
in addition to the density in \eqref{eq:posterior}. Proposal adaptation is restricted to warmup. CDF differences, latent transformations, and inverse quantiles are evaluated using stable log probability and survival probability calculations. Imputation uses joint posterior draws of the parameters and corresponding retained latent values. Posterior summaries use all post-warmup parameter draws. A prespecified subset of joint draws may be used to generate completed datasets, with the same indices used for every rule. Initialization, proposal tuning, and other implementation details are specified in the accompanying R code.

Model checking uses fresh Gaussian trajectories generated from \eqref{eq:copula} at every posterior parameter draw and transformed using the margins for the assigned treatment. These replicate trajectories retain dependence within participants. Randomized probability integral transform (PIT) residuals on the uniform scale are constructed by comparing each retained score with its simulated predictive distribution, accounting for discreteness.\cite{dunn1996,dharma} Their distribution is examined against a uniform reference distribution, overall and by treatment and visit, and residuals are plotted against fitted values. The residuals can retain dependence within participants, and their predictive distributions are estimated using the same observations being checked. Retention that depends on earlier outcomes can also affect these marginal residual checks. The resulting residual tests are therefore exploratory. These checks describe agreement at retained positions and do not validate the assumptions for outcomes after the trigger.

\subsection{Analysis of completed datasets}
\label{sec:analysis}

The analysis of completed datasets is chosen to address the scientific question and is not restricted to BB regression. Here, a BB mixed model with a participant random intercept provides an established approach to analyzing longitudinal discrete and bounded outcomes.\cite{najera2019betabinomial} Within a completed dataset, write the outcomes as $Y_{ij}$ and specify
\begin{equation}
\begin{aligned}
h_i &\overset{\mathrm{ind}}{\sim} \Normal\left(0, \sigma_h^2\right),\\
\left.Y_{ij} \middle| h_i\right.
&\sim \operatorname{BB}\left(K, \pi_{ij}, \kappa\right),\\
\logit\left(\pi_{ij}\right)
&= \delta_j + \vartheta_j A_i + \bm x_i^\top\bm\lambda + h_i,
\qquad \kappa > 0,\quad \sigma_h^2 \geq 0.
\end{aligned}
\label{eq:analysis}
\end{equation}
Here, $\delta_j$ and $\vartheta_j$ are visit-specific intercepts and treatment coefficients, $\bm\lambda = \left(\lambda_1, \ldots, \lambda_p\right)^\top$ contains covariate coefficients common across visits and treatment arms, and $h_i$ is independent of treatment and the covariates. The outcomes are conditionally independent across visits given $h_i$. The BB distribution uses beta shape parameters $\kappa\pi_{ij}$ and $\kappa\left(1 - \pi_{ij}\right)$, so $\kappa$ is a precision parameter shared across visits and arms. Its conditional moments are
\begin{equation}
\begin{aligned}
\E\left(Y_{ij} \middle| h_i\right) &= K\pi_{ij},\\
\Var\left(Y_{ij} \middle| h_i\right)
&= K\pi_{ij}\left(1 - \pi_{ij}\right)
\frac{K + \kappa}{1 + \kappa}.
\end{aligned}
\end{equation}
This is a working model for the bounded outcome. Its random intercept belongs to the analysis of completed datasets and can imply a different dependence structure from the Gaussian copula used for imputation.

Treatment comparisons use standardized mean scores on the original outcome scale. For $\sigma_h > 0$, let $\bm\psi$ collect the visit-specific coefficients $\delta_j$ and $\vartheta_j$, the components of $\bm\lambda$, and the coordinates $\log\kappa$ and $\log\sigma_h$. All parameter covariance matrices and derivatives below use these coordinates. For treatment level $g \in \left\{0, 1\right\}$, define
\begin{equation}
\begin{aligned}
\mu_j\left(g; \bm\psi\right)
&= \frac{K}{n}\sum_{i=1}^{n}\int_{-\infty}^{\infty}
\expit\left(\delta_j + \vartheta_j g + \bm x_i^\top\bm\lambda + \sigma_h z\right)
\varphi\left(z\right)\,\mathrm{d}z,\\
\Delta_j\left(\bm\psi\right)
&= \mu_j\left(1; \bm\psi\right) - \mu_j\left(0; \bm\psi\right),
\end{aligned}
\label{eq:standardized-means}
\end{equation}
where $\varphi$ is the standard normal density and $\sigma_h$ is the random intercept SD. The integral averages over the model's random intercept distribution through $h = \sigma_h z$. The outer average uses the empirical distribution of the covariates across all $n$ included participants, assigning each participant's covariate vector to both treatment levels. Thus, both arms are compared in the same covariate distribution. This is regression standardization, also called parametric g-computation.\cite{snowden2011} The covariate values and their equal weights are held fixed for inference, so the target is defined for the empirical covariate distribution of the included participants. The standardized means extend continuously to $\sigma_h = 0$, where the integral reduces to the inverse-logit prediction without a random effect.

The primary summary is $\Delta_J$, the standardized mean difference at the final visit; $\mu_j\left(g; \bm\psi\right)$ and $\Delta_j$ describe the mean profiles and contrasts over follow-up. The score units are unchanged by standardization, and a negative difference indicates a lower expected outcome in the active arm. With a nonlinear link, evaluating predictions at a zero random intercept or at the mean covariate vector generally does not give the average in \eqref{eq:standardized-means}. The median anchor indexes the imputation distribution, whereas the analysis of completed datasets targets mean scores.

Pooling is performed separately for each rule, with the rule index suppressed below. Let $M \geq 2$ be the number of completed datasets selected for analysis. In dataset $m = 1, \ldots, M$, fit the BB model by maximum likelihood and let $\widehat{\bm\psi}^{\left(m\right)}$ and $\widehat{\bm\Sigma}^{\left(m\right)}$ denote the parameter estimate and its joint model-based covariance matrix. The following delta-method calculation assumes a regular interior fit. For any one of the scalar summaries $s_j\left(\bm\psi\right) \in \left\{\mu_j\left(0; \bm\psi\right), \mu_j\left(1; \bm\psi\right), \Delta_j\left(\bm\psi\right)\right\}$, calculate
\begin{equation}
\begin{aligned}
\widehat s_j^{\left(m\right)}
&= s_j\left(\widehat{\bm\psi}^{\left(m\right)}\right),\\
\bm d_j^{\left(m\right)}
&= \left.\nabla_{\bm\psi}s_j\left(\bm\psi\right)\right|_{\bm\psi = \widehat{\bm\psi}^{\left(m\right)}},\\
U_j^{\left(m\right)}
&= \left(\bm d_j^{\left(m\right)}\right)^\top
\widehat{\bm\Sigma}^{\left(m\right)}\bm d_j^{\left(m\right)}.
\end{aligned}
\label{eq:delta-variance}
\end{equation}
Here, $\nabla_{\bm\psi}$ denotes the gradient with respect to the fitted parameter coordinates. The delta-method variance includes uncertainty in the fixed coefficients and random intercept SD, together with their covariance. Although the mean does not directly depend on $\kappa$, its joint estimation is accounted for in the parameter covariance matrix. For a treatment difference, the gradient is the difference between the two arm-specific gradients, retaining the covariance between the estimated means. Standardization and variance calculation are carried out within each completed dataset before pooling.

Rubin pooling\cite{rubin1987} uses
\begin{equation}
\begin{aligned}
\overline s_j &= \frac{1}{M}\sum_{m=1}^{M}\widehat s_j^{\left(m\right)},
& \overline U_j &= \frac{1}{M}\sum_{m=1}^{M}U_j^{\left(m\right)},\\
B_j^{\mathrm{MI}} &= \frac{1}{M - 1}\sum_{m=1}^{M}
\left(\widehat s_j^{\left(m\right)} - \overline s_j\right)^2,
& T_j &= \overline U_j + \left(1 + \frac{1}{M}\right)B_j^{\mathrm{MI}}.
\end{aligned}
\label{eq:rubin}
\end{equation}
Thus, $\overline U_j$ is the average variance within completed datasets, $B_j^{\mathrm{MI}}$ is the variance between their estimates, and $T_j$ is the pooled variance. Using the usual Rubin approximation with large-sample inference within each completed dataset, the nominal 95\% interval for each mean or difference is
\begin{equation}
\overline s_j \pm t_{\nu_j, 0.975}\sqrt{T_j},
\qquad
\nu_j = \left(M - 1\right)
\left\{1 + \frac{\overline U_j}{\left(1 + 1/M\right)B_j^{\mathrm{MI}}}\right\}^{2},
\label{eq:pooled-interval}
\end{equation}
where $t_{\nu, q}$ is the $q$ quantile of a Student $t$ distribution with $\nu$ degrees of freedom. When $B_j^{\mathrm{MI}} = 0$, use the normal limit with variance $\overline U_j$. Estimates and intervals remain on the original outcome scale. Although the mean estimates respect the outcome bounds, these approximate intervals are not constrained to the parameter space.

The conventional finite-$M$ variance correction and degrees-of-freedom approximation assume independent imputations. Spacing the selected draws within chains can reduce serial dependence but does not establish independence; $M$ is not an effective sample size. Any remaining dependence makes these pooling formulas approximate, and Monte Carlo stability must be assessed separately from posterior convergence.

The imputation and analysis models impose different distributional and dependence assumptions, and reference borrowing adds further restrictions. For longitudinal continuous outcomes, conditions have been established under which controlled multiple imputation with Rubin pooling is approximately information anchored, preserving the proportion of information lost because of missing data across the primary and sensitivity analyses.\cite{cro2019anchored} These results do not establish information anchoring for the present combination of DK imputation and BB analysis. Moreover, Rubin's variance can differ from the estimator's repeated-sampling variance under reference-based imputation.\cite{bartlett2023} The nominal intervals are therefore a prespecified pooling convention whose coverage requires evaluation.

\section{Application to Beating the Blues data}
\label{sec:application}

\subsection{Population and observed scores}

The \texttt{BtheB} dataset in \texttt{HSAUR2} contains 100 patients from a trial of computerized cognitive behavioral therapy for anxiety and depression.\cite{hsaur2,proudfoot2003} The comparison is between BtheB and treatment as usual (TAU). The package identifies the repeated outcome as the Beck Depression Inventory~II (BDI-II), measured at baseline and at months~2, 3, 5, and 8. The score ranges from 0 to 63, with higher scores indicating more severe symptoms.

The analysis includes participants with at least one recorded post-baseline score. Three patients with no recorded follow-up are excluded, leaving 97 participants: 45 receiving TAU and 52 receiving BtheB. Of 388 scheduled post-baseline measurements, 280 are observed, and 108 are missing (27.8\%). Missingness is monotone, and all 97 included participants have a score at month~2. There are 52 complete post-baseline trajectories. By month~8, 20 of 45 TAU scores (44.4\%) and 25 of 52 BtheB scores (48.1\%) are missing.

\begin{table}[ht!]
\caption{Observed post-baseline BDI-II scores in the 97-participant analysis population.}
\label{tab:observed}
\centering\small
\sisetup{
table-align-text-pre = false,
table-align-text-post = false
}
\begin{tabular}{
l
S[table-format = 1.0]
S[table-format = 2.0]
S[table-format = 2.0]
@{\;}
S[
table-format = 2.1,
table-space-text-pre = {(},
table-space-text-post = {)}
]
S[table-format = 2.1]
@{\;}
S[
table-format = 2.1,
table-space-text-pre = {(},
table-space-text-post = {)}
]
S[table-format = 2.1]
}
\toprule
\textbf{Treatment}
& {\textbf{Month}}
& {\textbf{Observed $\bm n$}}
& \multicolumn{2}{c}{\textbf{Missing $\bm n$ (\%)}}
& \multicolumn{2}{c}{\textbf{Mean (SD)}}
& {\textbf{Median}}\\
\midrule
TAU & 2 & 45 & 0 & {(}0.0{)} & 19.5 & {(}11.1{)} & 20.0\\
 & 3 & 36 & 9 & {(}20.0{)} & 17.7 & {(}12.7{)} & 15.5\\
 & 5 & 29 & 16 & {(}35.6{)} & 16.3 & {(}12.8{)} & 19.0\\
 & 8 & 25 & 20 & {(}44.4{)} & 13.6 & {(}11.5{)} & 13.0\\
\addlinespace
BtheB & 2 & 52 & 0 & {(}0.0{)} & 14.7 & {(}10.1{)} & 12.5\\
 & 3 & 37 & 15 & {(}28.8{)} & 12.0 & {(}10.4{)} & 10.0\\
 & 5 & 29 & 23 & {(}44.2{)} & 9.2 & {(}8.0{)} & 8.0\\
 & 8 & 27 & 25 & {(}48.1{)} & 8.9 & {(}6.1{)} & 9.0\\
\bottomrule
\end{tabular}
\par\smallskip
\begin{minipage}{\linewidth}
\footnotesize
Missing percentages use 45 TAU and 52 BtheB participants as denominators. Summary statistics use recorded scores only; the contributing participants change over follow-up.
\end{minipage}
\end{table}

Recorded mean scores decline in both arms and are lower in BtheB at every visit (\maintabref{tab:observed}). These are comparisons among the available observations, whose composition changes over follow-up. Eighteen recorded scores are zero (6.4\%); none equals the upper bound. The known upper bound is nevertheless retained as part of the outcome model.

\subsection{Imputation model and assessment}
\label{sec:application-computation}

The imputation model uses $K = 63$, $J = 4$, and visit times $\left(t_1, t_2, t_3, t_4\right) = \left(2, 3, 5, 8\right)$ measured in months. Treatment is coded as $A_i = 0$ for TAU and $A_i = 1$ for BtheB. The sole covariate is the fully observed baseline BDI-II score $Y_{i0}$, centered and scaled as $B_i = \left(Y_{i0} - c_B\right)/s_B$, where $c_B = 23.1546$ is the mean among included participants and $s_B = 10$. Thus, $p = 1$ and $\bm x_i = \left(B_i\right)$, with visit-specific baseline coefficients $\gamma_{j1}$ in the imputation model.

Clinical event dates are unavailable, so the operational trigger is $D_i = 1 + \max \mathcal R_i$. Every recorded score is therefore retained, giving $\mathcal O_i = \mathcal R_i$, and reference borrowing applies only to the missing final tail. Under the observed monotone pattern, $D_i$ is the first missing visit for an incomplete trajectory; a complete trajectory has $D_i = J + 1$.

The prior hyperparameters in \eqref{eq:priors} are set to
\begin{equation}
\begin{gathered}
\mu_\alpha = \mu_\beta = \mu_{\gamma 1} = \mu_\phi = 0,\\
\sigma_\alpha = \sigma_\beta = \sigma_{\gamma 1} = 5,
\qquad \sigma_\phi = 2,
\qquad a_\rho = b_\rho = 1.
\end{gathered}
\label{eq:application-priors}
\end{equation}
These choices give independent $\Normal\left(0, 5^2\right)$ priors for $\alpha_j$, $\beta_j$, and $\gamma_{j1}$, a $\Normal\left(0, 2^2\right)$ prior for $\log\phi_j$, and a uniform prior on $\rho \in \left(0, 1\right)$. The prior for $\gamma_{j1}$ refers to a ten-point difference in baseline BDI-II, and the correlation prior refers to the latent correlation for a one-month separation. The same hyperparameter choices are used in the simulation studies.

Four parallel chains each comprise 5,000 warmup iterations and 10,000 retained iterations. All 40,000 post-warmup draws are used for posterior summaries, predictive checks, and imputation.

Convergence and sampling efficiency are assessed using rank-normalized split R-hat ($\widehat R$) and bulk and tail effective sample sizes (ESSs).\cite{vehtari2021} Across the 17 model parameters, the largest $\widehat R$ was 1.005, and the smallest bulk and tail ESSs were 690 and 1,344, respectively.

\supptabref{tab:dk-parameters} in the \supportingref{supp:parameters} summarizes the fitted imputation model.

Residual diagnostics use \texttt{DHARMa},\cite{dharma} with all 40,000 replicated datasets and the 280 recorded post-baseline scores. The quantile plot follows the uniform reference line reasonably closely, with modest departures (\mainfigref{fig:residuals}a). The exploratory tests give $p = 0.147$ for uniformity, $p = 0.306$ for dispersion, $p = 0.496$ for zero inflation, and $p = 1.000$ for outliers. These exploratory tests do not indicate an overall discrepancy in the features assessed.

Residual distributions vary across treatment and visit, with BtheB residuals more concentrated around the center at month~8 (\mainfigref{fig:residuals}b). Residuals against fitted values also show local trends, particularly in BtheB at months~3 and 8 (\suppfigref{fig:residuals-fitted} in the \supportingref{supp:residuals}). These checks assess fit at recorded visits and do not validate the assumptions for unobserved scores.

\begin{figure}[p]
\centering
\captionsetup{font=small}
\begin{subfigure}{0.54\textwidth}
\centering
\includegraphics[width=\linewidth]{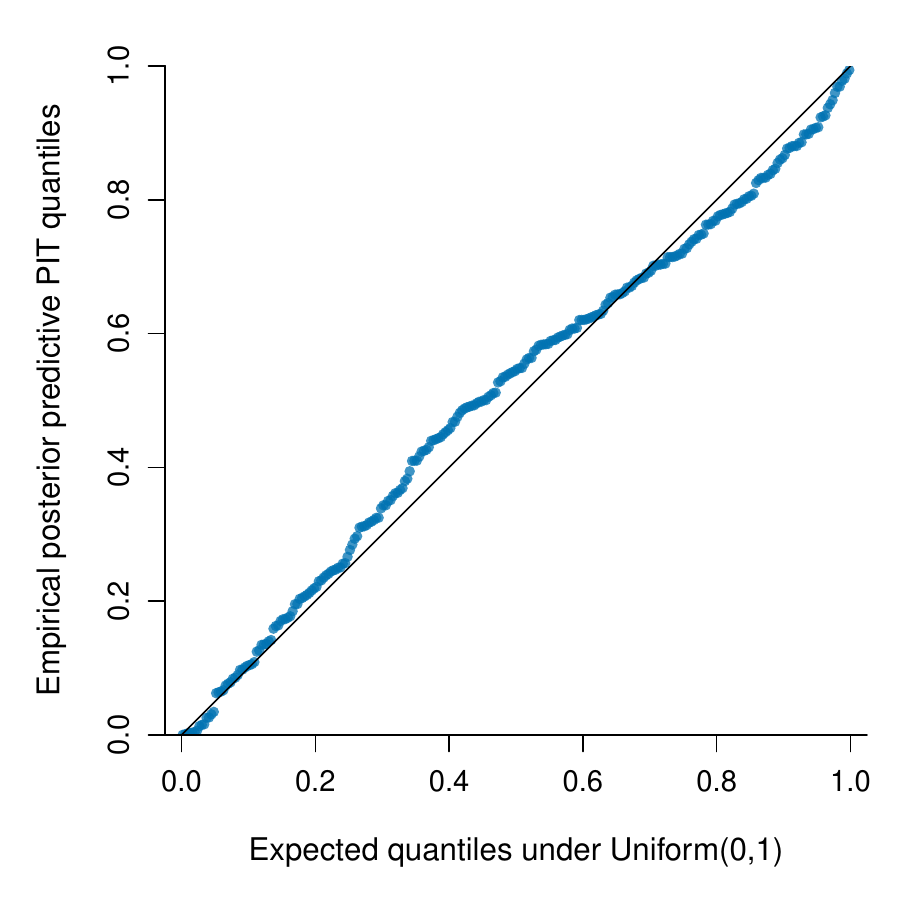}
\caption{Overall quantile plot.}
\end{subfigure}

\medskip
\begin{subfigure}{0.86\textwidth}
\centering
\includegraphics[width=\linewidth]{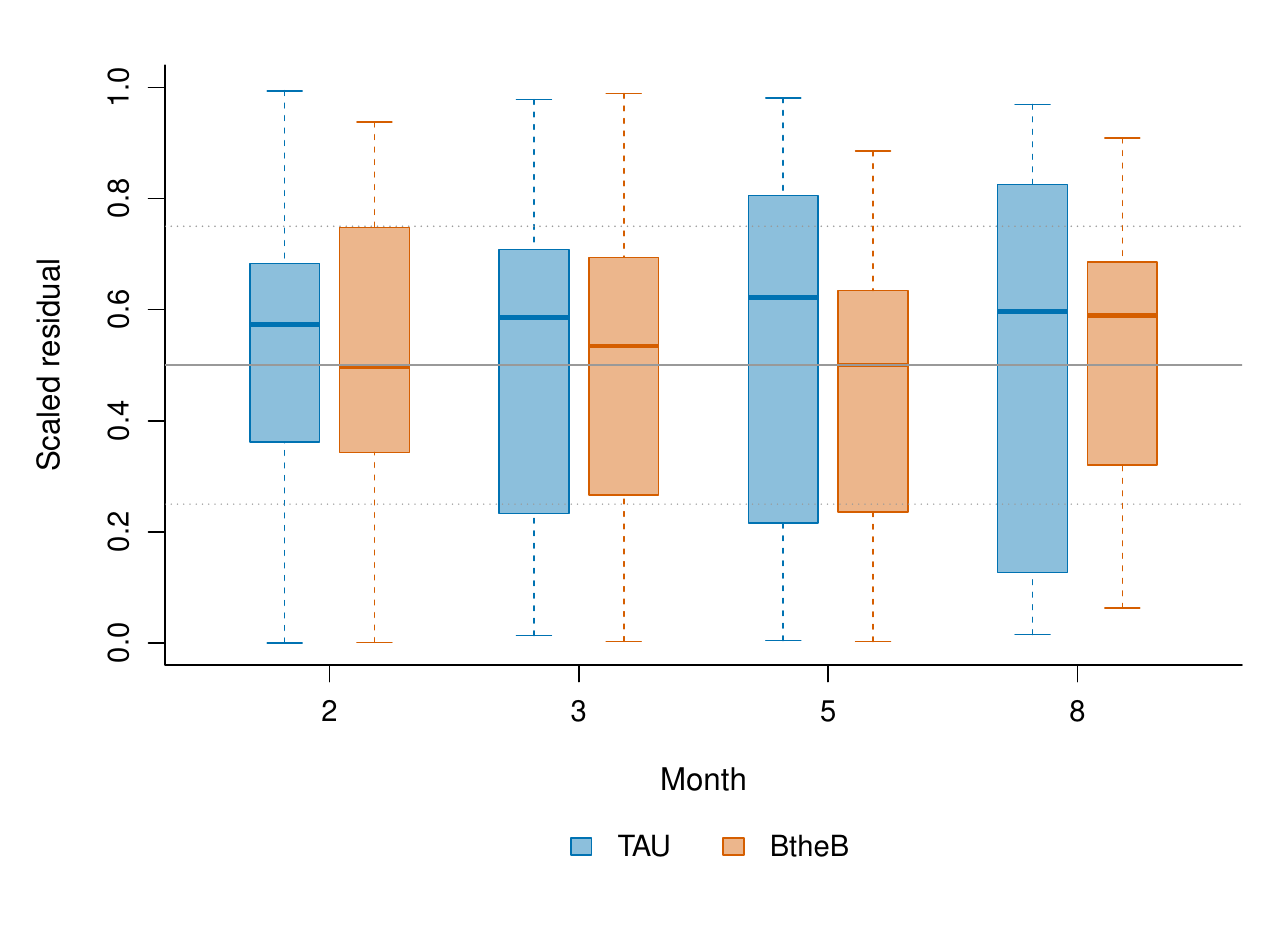}
\caption{Residual distributions by treatment and visit.}
\end{subfigure}
\caption{DHARMa randomized predictive residuals at the 280 recorded positions, using all 40,000 replicated datasets. Panel (a) includes the uniform reference line. Horizontal lines in panel (b) mark 0.25, 0.50, and 0.75.}
\label{fig:residuals}
\end{figure}

\subsection{Imputations and treatment comparisons}
\label{sec:application-analysis}

The application uses MAR, halfway borrowing, and J2R, with borrowing weights $\omega = 0$, $\omega = 1/2$, and $\omega = 1$, respectively. One completed dataset is generated per posterior draw and rule, giving 40,000 completed datasets per rule. Each dataset preserves all 280 recorded scores and fills the 108 missing positions with integers between 0 and 63. With the paired latent draws, TAU imputations are identical across rules. Only the missing BtheB final tails are affected by reference borrowing. Because the observed pattern is monotone, no intermittent gaps require separate handling in this application.

For the treatment analysis, $M = 100$ completed datasets are selected per rule, using 25 evenly spaced post-warmup iterations from each of the four chains. The same draw indices are used for every rule. This limits the number of BB fits to 300 while retaining all 40,000 posterior draws for the imputation model summaries and residual diagnostics. All 300 BB fits passed the convergence and numerical checks, and all 100 selected datasets per rule contributed to pooling.

Each selected dataset is analyzed using the BB random intercept model in \eqref{eq:analysis}, with $\bm x_i = \left(B_i\right)$ and a common baseline coefficient $\lambda_1$. Models are fitted by maximum likelihood using \texttt{glmmTMB}, with a Laplace approximation to the likelihood integral over random effects.\cite{brooks2017} For the standardized means in \eqref{eq:standardized-means}, a separate Gauss-Hermite quadrature integrates the inverse-logit predictions over the fitted normal random intercept distribution. Quadrature accuracy is checked by increasing the number of nodes. Predictions are averaged over all 97 baseline scores, using the same empirical distribution for both treatment levels. Thus, the reported means average over baseline severity and the random intercept distribution; they are not predictions at the centered baseline value.

The primary comparison is the standardized mean difference at month~8, $\Delta_4$, expressed as BtheB minus TAU. Means and differences are computed within each completed dataset, with delta-method variances based on the joint covariance of the fixed coefficients and log random intercept SD. These estimates and variances are pooled separately within each rule using \eqref{eq:rubin} and \eqref{eq:pooled-interval}. The empirical distribution of the 97 observed baseline scores is treated as fixed for inference. Separate analyses using the selected imputations from each chain gave negative differences at month~8 under all three rules, with confidence intervals including zero in every case.

\maintabref{tab:bb-analysis} presents the standardized mean profiles and treatment differences. All reported treatment differences favor BtheB numerically, but every confidence interval includes zero. Stronger reference borrowing moves the estimated difference at the final visit toward zero without changing this conclusion.

Small differences in the estimated TAU means across rules arise because each BB fit jointly re-estimates the common baseline coefficient, precision, and random intercept variance using both arms. The same feature allows small changes at month~2 even though all scores at that visit are observed. Identical imputations for a subset of observations can therefore yield different standardized estimates from the fitted analysis model.

\begin{table}[tbp]
\caption{Standardized mean BDI-II scores and treatment differences after reference-based imputation.}
\label{tab:bb-analysis}
\centering
\begingroup\small
\setlength{\tabcolsep}{3pt}
\sisetup{table-align-text-pre = false, table-align-text-post = false,
  group-digits = false, output-decimal-marker = {.}, round-mode = none}
\begin{tabular*}{\linewidth}{@{\extracolsep{\fill}}lc
S[table-format = 2.2]
@{\hspace{\fontdimen2\font}\extracolsep{0pt}}
S[table-format = 2.2, table-space-text-pre = {(}, table-space-text-post = {,}]
@{\hspace{\fontdimen2\font}}
S[table-format = 2.2, table-space-text-post = {)}]
@{\hspace{1em}\extracolsep{\fill}}
S[table-format = 2.2]
@{\hspace{\fontdimen2\font}\extracolsep{0pt}}
S[table-format = 2.2, table-space-text-pre = {(}, table-space-text-post = {,}]
@{\hspace{\fontdimen2\font}}
S[table-format = 2.2, table-space-text-post = {)}]
@{\hspace{1em}\extracolsep{\fill}}
S[table-format = -1.2]
@{\hspace{\fontdimen2\font}\extracolsep{0pt}}
S[table-format = -1.2, table-space-text-pre = {(}, table-space-text-post = {,}]
@{\hspace{\fontdimen2\font}}
S[table-format = 1.2, table-space-text-post = {)}]
@{}
}
\toprule
& & \multicolumn{6}{c}{\textbf{Standardized mean}} & \multicolumn{3}{c}{\textbf{Mean difference}} \\
\cmidrule(lr){3-8}\cmidrule(l){9-11}
\textbf{Rule} & \textbf{Month} & \multicolumn{3}{c}{\textbf{TAU}} & \multicolumn{3}{c}{\textbf{BtheB}} & \multicolumn{3}{c}{\textbf{BtheB minus TAU}} \\
\midrule
MAR & 2 & 18.94 & {(}15.87{,} & 22.00{)} & 15.28 & {(}12.65{,} & 17.91{)} & -3.65 & {(}-7.67{,} & 0.36{)} \\
 & 3 & 17.79 & {(}14.75{,} & 20.84{)} & 14.14 & {(}11.35{,} & 16.93{)} & -3.65 & {(}-7.72{,} & 0.42{)} \\
 & 5 & 14.92 & {(}11.72{,} & 18.12{)} & 12.39 & {(}9.04{,} & 15.74{)} & -2.52 & {(}-6.96{,} & 1.91{)} \\
 & 8 & 12.33 & {(}8.77{,} & 15.88{)} & 10.60 & {(}7.28{,} & 13.92{)} & -1.72 & {(}-6.57{,} & 3.12{)} \\
\addlinespace
Halfway & 2 & 18.94 & {(}15.86{,} & 22.03{)} & 15.29 & {(}12.65{,} & 17.94{)} & -3.65 & {(}-7.68{,} & 0.38{)} \\
 & 3 & 17.80 & {(}14.75{,} & 20.86{)} & 14.72 & {(}11.92{,} & 17.52{)} & -3.09 & {(}-7.17{,} & 0.99{)} \\
 & 5 & 14.94 & {(}11.72{,} & 18.15{)} & 12.89 & {(}9.76{,} & 16.02{)} & -2.04 & {(}-6.27{,} & 2.18{)} \\
 & 8 & 12.34 & {(}8.77{,} & 15.91{)} & 10.91 & {(}7.75{,} & 14.06{)} & -1.44 & {(}-5.87{,} & 3.00{)} \\
\addlinespace
J2R & 2 & 18.96 & {(}15.86{,} & 22.06{)} & 15.31 & {(}12.64{,} & 17.97{)} & -3.65 & {(}-7.72{,} & 0.41{)} \\
 & 3 & 17.82 & {(}14.74{,} & 20.90{)} & 15.35 & {(}12.41{,} & 18.29{)} & -2.47 & {(}-6.65{,} & 1.71{)} \\
 & 5 & 14.96 & {(}11.73{,} & 18.19{)} & 13.44 & {(}10.17{,} & 16.72{)} & -1.51 & {(}-5.80{,} & 2.77{)} \\
 & 8 & 12.37 & {(}8.78{,} & 15.95{)} & 11.27 & {(}7.90{,} & 14.65{)} & -1.09 & {(}-5.33{,} & 3.15{)} \\
\bottomrule
\end{tabular*}
\endgroup
\par\smallskip
\begin{minipage}{\linewidth}\footnotesize
Predictions from the BB random intercept model are integrated over the fitted random intercept distribution and averaged over the baseline scores of all 97 included participants, using the same distribution for both arms. Estimates and delta-method variances are pooled separately within each rule using 100 completed datasets selected at evenly spaced iterations within each chain. The same draw indices are used across rules. Intervals are 95\% confidence intervals obtained using Rubin pooling with large-sample inference within each completed dataset. The primary comparison is the mean difference at month 8; negative differences favor BtheB.
\end{minipage}
\end{table}

\FloatBarrier
\section{Simulation studies}
\label{sec:simulation}

\subsection{Recovery of model parameters}
\label{sec:simulation-recovery}

The first study evaluates estimation of the DK Gaussian copula model using fully observed data generated from \eqref{eq:copula}. The design uses $K = 63$, four visits at months~2, 3, 5, and 8, and sample sizes $n = 100$ and $n = 200$, with equal allocation to the two treatment arms. For each sample size, baseline scores are sampled with replacement from the 97 included BtheB participants and centered and scaled as in the application. The covariate values and treatment allocation are then held fixed across replicates and shared between the marginal settings.

The first marginal setting uses generating parameters informed by a preliminary application fit, with values listed in \maintabref{tab:dk-recovery-full-application}. These values remain fixed throughout the study. The second setting subtracts 2 from every visit intercept while retaining the treatment coefficients, baseline coefficients, shapes, and correlation, increasing concentration at zero (\maintabref{tab:dk-recovery-full-floor}). Each dataset is fitted using the generating model and the prior hyperparameters in \eqref{eq:application-priors}.

The study comprised 300 replicates per scenario, giving 1,200 replicates in total. Each fit uses four chains with 5,000 warmup and 10,000 post-warmup iterations per chain. All 40,000 post-warmup draws contribute to inference. Evaluation concerns the 17 parameters $\alpha_j$, $\beta_j$, $\gamma_{j1}$, and $\phi_j$, $j = 1, \ldots, 4$, and $\rho$. Bias and root mean squared error (RMSE) are calculated for posterior means; coverage and mean interval width are calculated for 95\% equal-tailed credible intervals. Shape and correlation summaries are computed after transforming each draw to its natural scale. All 1,200 fits were completed and contributed to the performance summaries.

\maintabref{tab:dk-recovery-full-application} and \maintabref{tab:dk-recovery-full-floor} summarize parameter recovery. Bias was generally small, and empirical coverage ranged from 92.0\% to 97.3\% across the regression coefficients, shapes, and correlation, remaining close to the nominal 95\% level. Increasing the sample size reduced RMSE and mean interval width for every parameter in both settings. Greater concentration at zero increased uncertainty in the regression coefficients; modest negative bias in the intercepts was smaller at $n=200$ than at $n=100$.

Across all model parameters and replicates, $\widehat R$ was below 1.05. The minimum ESS, taken over both bulk and tail values across all parameters and replicates, was 243.

\afterpage{\begin{landscape}
\begin{table}[p]
\centering\small
\caption{Simulation: parameter recovery from fully observed data with margins resembling those in the application.}
\label{tab:dk-recovery-full-application}
\setlength{\tabcolsep}{5pt}
\renewcommand{\arraystretch}{1.12}
\sisetup{table-align-text-pre = false, table-align-text-post = false, group-digits = false, output-decimal-marker = {.}, round-mode = none}
\begin{tabular*}{\linewidth}{@{\extracolsep{\fill}}l S[table-format = -1.3] S[table-format = -1.3] S[table-format = 1.3] S[table-format = 2.1] @{\extracolsep{0pt}\;} S[table-format = 1.1, table-space-text-pre = {(}, table-space-text-post = {)}] @{\extracolsep{\fill}\hspace{2\tabcolsep}} S[table-format = 1.3] S[table-format = -1.3] S[table-format = 1.3] S[table-format = 2.1] @{\extracolsep{0pt}\;} S[table-format = 1.1, table-space-text-pre = {(}, table-space-text-post = {)}] @{\extracolsep{\fill}\hspace{2\tabcolsep}} S[table-format = 1.3]@{}}
\toprule
 & & \multicolumn{5}{c}{$\boldsymbol{n = 100}$} & \multicolumn{5}{c}{$\boldsymbol{n = 200}$} \\
\cmidrule(lr){3-7}\cmidrule(lr){8-12}
\textbf{Parameter} & {\textbf{Truth}} & {\textbf{Bias}} & {\textbf{RMSE}} & \multicolumn{2}{c}{\shortstack{\textbf{Coverage (\%)}\\\textbf{(MCSE)}}} & {\textbf{Mean width}} & {\textbf{Bias}} & {\textbf{RMSE}} & \multicolumn{2}{c}{\shortstack{\textbf{Coverage (\%)}\\\textbf{(MCSE)}}} & {\textbf{Mean width}} \\
\midrule
$\alpha_{1}$ & -0.950 & -0.018 & 0.125 & 93.7 & {(}1.4{)} & 0.483 & -0.013 & 0.084 & 95.0 & {(}1.3{)} & 0.332 \\
$\alpha_{2}$ & -1.110 & -0.026 & 0.152 & 93.3 & {(}1.4{)} & 0.583 & -0.014 & 0.095 & 96.3 & {(}1.1{)} & 0.399 \\
$\alpha_{3}$ & -1.480 & -0.030 & 0.160 & 94.3 & {(}1.3{)} & 0.644 & -0.009 & 0.109 & 96.3 & {(}1.1{)} & 0.440 \\
$\alpha_{4}$ & -1.740 & -0.023 & 0.170 & 96.0 & {(}1.1{)} & 0.660 & -0.012 & 0.107 & 95.0 & {(}1.3{)} & 0.454 \\
\addlinespace[3pt]
$\beta_{1}$ & -0.380 & -0.009 & 0.173 & 95.7 & {(}1.2{)} & 0.707 & 0.013 & 0.119 & 94.7 & {(}1.3{)} & 0.479 \\
$\beta_{2}$ & -0.430 & -0.009 & 0.208 & 94.3 & {(}1.3{)} & 0.847 & 0.010 & 0.151 & 93.3 & {(}1.4{)} & 0.569 \\
$\beta_{3}$ & -0.280 & -0.003 & 0.228 & 94.0 & {(}1.4{)} & 0.910 & -0.001 & 0.161 & 94.3 & {(}1.3{)} & 0.610 \\
$\beta_{4}$ & -0.160 & -0.002 & 0.233 & 95.0 & {(}1.3{)} & 0.930 & 0.004 & 0.148 & 95.3 & {(}1.2{)} & 0.628 \\
\addlinespace[3pt]
$\gamma_{11}$ & 0.440 & -0.001 & 0.076 & 96.0 & {(}1.1{)} & 0.311 & 0.005 & 0.051 & 96.0 & {(}1.1{)} & 0.202 \\
$\gamma_{21}$ & 0.600 & 0.002 & 0.086 & 97.3 & {(}0.9{)} & 0.367 & 0.003 & 0.062 & 95.7 & {(}1.2{)} & 0.240 \\
$\gamma_{31}$ & 0.630 & -0.005 & 0.094 & 96.7 & {(}1.0{)} & 0.396 & 0.002 & 0.070 & 92.7 & {(}1.5{)} & 0.259 \\
$\gamma_{41}$ & 0.520 & -0.003 & 0.107 & 94.3 & {(}1.3{)} & 0.409 & 0.000 & 0.070 & 93.7 & {(}1.4{)} & 0.264 \\
\addlinespace[3pt]
$\phi_{1}$ & 2.320 & 0.009 & 0.169 & 96.3 & {(}1.1{)} & 0.705 & 0.003 & 0.120 & 96.0 & {(}1.1{)} & 0.494 \\
$\phi_{2}$ & 1.990 & 0.009 & 0.172 & 95.7 & {(}1.2{)} & 0.646 & -0.009 & 0.108 & 96.0 & {(}1.1{)} & 0.451 \\
$\phi_{3}$ & 2.030 & 0.003 & 0.182 & 94.0 & {(}1.4{)} & 0.680 & -0.016 & 0.116 & 96.0 & {(}1.1{)} & 0.474 \\
$\phi_{4}$ & 2.150 & 0.002 & 0.176 & 97.0 & {(}1.0{)} & 0.709 & -0.011 & 0.119 & 96.0 & {(}1.1{)} & 0.495 \\
\addlinespace[3pt]
$\rho$ & 0.780 & -0.005 & 0.030 & 92.7 & {(}1.5{)} & 0.110 & -0.001 & 0.020 & 94.0 & {(}1.4{)} & 0.075 \\
\bottomrule
\end{tabular*}
\par\smallskip
\begin{minipage}{\linewidth}\footnotesize
All 300 replicates contributed at each sample size. Bias and root mean squared error (RMSE) refer to posterior means. Coverage and mean width refer to 95\% equal-tailed credible intervals. Parentheses give the simulation Monte Carlo standard error (MCSE) of coverage in percentage points. Each fit uses all 40,000 post-warmup draws. Shape and correlation are summarized on their natural scales. Visits $j = 1,\ldots,4$ correspond to months 2, 3, 5, 8.
\end{minipage}
\end{table}

\begin{table}[p]
\centering\small
\caption{Simulation: parameter recovery from fully observed data with greater concentration at zero.}
\label{tab:dk-recovery-full-floor}
\setlength{\tabcolsep}{5pt}
\renewcommand{\arraystretch}{1.12}
\sisetup{table-align-text-pre = false, table-align-text-post = false, group-digits = false, output-decimal-marker = {.}, round-mode = none}
\begin{tabular*}{\linewidth}{@{\extracolsep{\fill}}l S[table-format = -1.3] S[table-format = -1.3] S[table-format = 1.3] S[table-format = 2.1] @{\extracolsep{0pt}\;} S[table-format = 1.1, table-space-text-pre = {(}, table-space-text-post = {)}] @{\extracolsep{\fill}\hspace{2\tabcolsep}} S[table-format = 1.3] S[table-format = -1.3] S[table-format = 1.3] S[table-format = 2.1] @{\extracolsep{0pt}\;} S[table-format = 1.1, table-space-text-pre = {(}, table-space-text-post = {)}] @{\extracolsep{\fill}\hspace{2\tabcolsep}} S[table-format = 1.3]@{}}
\toprule
 & & \multicolumn{5}{c}{$\boldsymbol{n = 100}$} & \multicolumn{5}{c}{$\boldsymbol{n = 200}$} \\
\cmidrule(lr){3-7}\cmidrule(lr){8-12}
\textbf{Parameter} & {\textbf{Truth}} & {\textbf{Bias}} & {\textbf{RMSE}} & \multicolumn{2}{c}{\shortstack{\textbf{Coverage (\%)}\\\textbf{(MCSE)}}} & {\textbf{Mean width}} & {\textbf{Bias}} & {\textbf{RMSE}} & \multicolumn{2}{c}{\shortstack{\textbf{Coverage (\%)}\\\textbf{(MCSE)}}} & {\textbf{Mean width}} \\
\midrule
$\alpha_{1}$ & -2.950 & -0.086 & 0.243 & 95.7 & {(}1.2{)} & 0.934 & -0.043 & 0.162 & 94.7 & {(}1.3{)} & 0.625 \\
$\alpha_{2}$ & -3.110 & -0.113 & 0.289 & 97.0 & {(}1.0{)} & 1.143 & -0.057 & 0.189 & 96.3 & {(}1.1{)} & 0.763 \\
$\alpha_{3}$ & -3.480 & -0.129 & 0.345 & 95.7 & {(}1.2{)} & 1.305 & -0.061 & 0.235 & 94.0 & {(}1.4{)} & 0.853 \\
$\alpha_{4}$ & -3.740 & -0.119 & 0.369 & 94.7 & {(}1.3{)} & 1.353 & -0.038 & 0.236 & 94.0 & {(}1.4{)} & 0.896 \\
\addlinespace[3pt]
$\beta_{1}$ & -0.380 & 0.025 & 0.336 & 95.0 & {(}1.3{)} & 1.353 & 0.002 & 0.245 & 93.3 & {(}1.4{)} & 0.900 \\
$\beta_{2}$ & -0.430 & 0.035 & 0.391 & 95.0 & {(}1.3{)} & 1.640 & 0.008 & 0.259 & 94.7 & {(}1.3{)} & 1.076 \\
$\beta_{3}$ & -0.280 & 0.025 & 0.450 & 95.7 & {(}1.2{)} & 1.804 & -0.004 & 0.298 & 94.0 & {(}1.4{)} & 1.162 \\
$\beta_{4}$ & -0.160 & -0.015 & 0.465 & 94.7 & {(}1.3{)} & 1.865 & -0.028 & 0.304 & 93.7 & {(}1.4{)} & 1.219 \\
\addlinespace[3pt]
$\gamma_{11}$ & 0.440 & -0.002 & 0.152 & 94.3 & {(}1.3{)} & 0.584 & -0.003 & 0.095 & 95.0 & {(}1.3{)} & 0.375 \\
$\gamma_{21}$ & 0.600 & -0.007 & 0.189 & 93.3 & {(}1.4{)} & 0.693 & 0.000 & 0.113 & 95.0 & {(}1.3{)} & 0.449 \\
$\gamma_{31}$ & 0.630 & 0.006 & 0.190 & 96.3 & {(}1.1{)} & 0.770 & 0.002 & 0.131 & 92.0 & {(}1.6{)} & 0.488 \\
$\gamma_{41}$ & 0.520 & 0.006 & 0.197 & 96.3 & {(}1.1{)} & 0.807 & -0.013 & 0.129 & 95.3 & {(}1.2{)} & 0.512 \\
\addlinespace[3pt]
$\phi_{1}$ & 2.320 & -0.024 & 0.198 & 94.0 & {(}1.4{)} & 0.796 & 0.006 & 0.141 & 94.3 & {(}1.3{)} & 0.556 \\
$\phi_{2}$ & 1.990 & -0.017 & 0.181 & 95.7 & {(}1.2{)} & 0.739 & -0.009 & 0.131 & 94.0 & {(}1.4{)} & 0.515 \\
$\phi_{3}$ & 2.030 & -0.035 & 0.212 & 93.0 & {(}1.5{)} & 0.801 & -0.004 & 0.147 & 94.0 & {(}1.4{)} & 0.564 \\
$\phi_{4}$ & 2.150 & -0.025 & 0.218 & 95.0 & {(}1.3{)} & 0.873 & -0.030 & 0.153 & 93.0 & {(}1.5{)} & 0.603 \\
\addlinespace[3pt]
$\rho$ & 0.780 & -0.003 & 0.034 & 95.0 & {(}1.3{)} & 0.124 & -0.004 & 0.023 & 94.0 & {(}1.4{)} & 0.087 \\
\bottomrule
\end{tabular*}
\par\smallskip
\begin{minipage}{\linewidth}\footnotesize
All 300 replicates contributed at each sample size. Bias and root mean squared error (RMSE) refer to posterior means. Coverage and mean width refer to 95\% equal-tailed credible intervals. Parentheses give the simulation Monte Carlo standard error (MCSE) of coverage in percentage points. Each fit uses all 40,000 post-warmup draws. Shape and correlation are summarized on their natural scales. Visits $j = 1,\ldots,4$ correspond to months 2, 3, 5, 8.
\end{minipage}
\end{table}
\end{landscape}}

\subsection{Imputation and analysis of mean scores}
\label{sec:simulation-imputation}

The second study comprised 100 replicates with $n = 100$, using the generating parameters in \maintabref{tab:dk-recovery-full-application} and the fixed design for $n = 100$ described in \mainsecref{sec:simulation-recovery}. All month~2 scores are observed. Subsequently, dropout at visit $j$, conditional on remaining observed through visit $j - 1$, has probability
\begin{equation}
\Pr\left(D_i = j \middle| D_i \geq j, B_i, Y_{i,j-1}\right)
= \expit\left\{\upsilon_j + \log\left(1.25\right)B_i
+ \log\left(1.5\right)\frac{Y_{i,j-1} - 20}{10}\right\},
\qquad j = 2, 3, 4.
\label{eq:simulation-dropout}
\end{equation}
The intercepts $\upsilon_j$ are calibrated using 200,000 independent trajectories to cumulative missingness of 25\%, 40\%, and 45\% at months~3, 5, and 8, respectively, pooled across arms. Scores from $D_i$ onward are withheld; there are no intermittent gaps.

Within each replicate, MAR trajectories and triggers are generated first. Halfway and J2R versions replace scores in the active arm from the trigger onward using \eqref{eq:rule} and the same latent Gaussian values. Thus, all generating assumptions share the retained outcomes and require one DK fit per replicate. Priors and chain lengths match the first study. Posterior summaries use all 40,000 post-warmup draws; imputation uses $M = 50$ joint posterior draws selected at evenly spaced iterations within chains and allocated as evenly as possible across the four chains. The selected draws and missing latent values are shared across rules. The resulting datasets are analyzed and pooled as in \mainsecref{sec:analysis}.

Targets are the active mean at month~8 and the active minus reference mean difference, standardized to all 100 fixed baseline values. Under MAR, the true standardized means in both arms are calculated exactly by evaluating \eqref{eq:dk-mean} at the generating parameter values and averaging over those baseline values. For the truth calculation, let $\widetilde Y_{iJ}^{\left(\omega\right)}$ denote a generated final score before deletion under generating assumption $\omega$, and let $\mu_{J,\mathrm{true}}^{\left(\omega\right)}\left(1\right)$ denote the corresponding true standardized mean in the active arm. Then
\begin{equation}
\mu_{J,\mathrm{true}}^{\left(\omega\right)}\left(1\right)
= \mu_{J,\mathrm{true}}^{\left(0\right)}\left(1\right)
+ \frac{1}{n}\sum_{i=1}^{n}
\E\left[
\widetilde Y_{iJ}^{\left(\omega\right)}
- \widetilde Y_{iJ}^{\left(0\right)}
\middle| A_i = 1, \bm x_i
\right].
\label{eq:simulation-truth}
\end{equation}
Here, the paired scores use the same latent trajectory and dropout trigger. The expectation averages over outcome histories and dropout patterns at the fixed generating parameter values. The second term in \eqref{eq:simulation-truth} is estimated using at least two million trajectories generated independently of the study replicates. For each trajectory, dropout is averaged over using the probabilities from \eqref{eq:simulation-dropout}. The target Monte Carlo standard error (MCSE) for the estimated change is below 0.005 score units. Subtracting the unchanged reference mean gives the true treatment difference.

Benchmarks are BB fits before deletion and, under MAR, standardized posterior means from the DK model fitted to incomplete data. Every imputation rule is evaluated against all three truths; mismatched rules assess sensitivity rather than interval calibration. Performance measures are bias, RMSE, coverage of nominal 95\% intervals, and mean interval width. MCSEs are reported for coverage.

\maintabref{tab:imputation-simulation} summarizes the active mean. Stronger borrowing increases the estimated mean in the expected direction, but the matching rule does not consistently give the smallest absolute bias. Modest downward bias is present in the BB estimates even before deletion, with empirical coverage below nominal; the direct DK estimate under MAR is closer to its target. Under matching imputation, empirical coverage was 95-98\% for the active mean and 98-99\% for treatment differences (\maintabref{tab:imputation-contrasts}). Under halfway borrowing and J2R, the treatment differences had smaller absolute biases than the active means, partly reflecting cancellation between arms. The expected direction of change under borrowing therefore does not by itself establish accurate estimation or interval calibration.

\begin{table}[tbp]
\caption{Standardized mean in the active arm at month~8 in the imputation simulation ($n=100$).}
\label{tab:imputation-simulation}
\centering\small
\begingroup
\setlength{\tabcolsep}{3pt}
\renewcommand{\arraystretch}{1.08}
\sisetup{group-digits = false, round-mode = none, table-align-text-pre = false, table-align-text-post = false}
\begin{tabular}{
@{}l
@{\hspace{12pt}}
l
@{\hspace{12pt}}
S[table-format = 2.2]
S[table-format = -1.2]
S[table-format = 1.2]
S[table-format = 2.1]
@{\hspace{\fontdimen2\font}}
S[table-format = 1.1, table-space-text-pre = {(}, table-space-text-post = {)}]
S[table-format = 1.2]
@{}}
\toprule
\textbf{Analysis} & \textbf{Imputation rule} & {\textbf{Mean}} & {\textbf{Bias}} & {\textbf{RMSE}} & \multicolumn{2}{c}{\shortstack{\textbf{Coverage (\%)}\\\textbf{(MCSE)}}} & {\textbf{Width}} \\
\midrule
\multicolumn{8}{@{}l}{\textbf{Generating assumption: MAR; true mean $\bm{10.82}$}} \\
\addlinespace[2pt]
Complete data, BB & --- & 10.54 & -0.28 & 1.23 & 89.0 & {(}3.1{)} & 4.59 \\
Observed data, DK & --- & 10.90 & 0.08 & 1.61 & 95.0 & {(}2.2{)} & 6.73 \\
DK imputation, BB & MAR & 10.54 & -0.28 & 1.58 & 95.0 & {(}2.2{)} & 6.48 \\
 & Halfway & 10.70 & -0.12 & 1.39 & 96.0 & {(}2.0{)} & 6.24 \\
 & J2R & 10.93 & 0.11 & 1.36 & 99.0 & {(}1.0{)} & 6.71 \\
\addlinespace[5pt]
\multicolumn{8}{@{}l}{\textbf{Generating assumption: Halfway; true mean $\bm{11.09}$}} \\
\addlinespace[2pt]
Complete data, BB & --- & 10.79 & -0.30 & 1.24 & 91.0 & {(}2.9{)} & 4.65 \\
DK imputation, BB & MAR & 10.54 & -0.55 & 1.65 & 91.0 & {(}2.9{)} & 6.48 \\
 & Halfway & 10.70 & -0.40 & 1.44 & 96.0 & {(}2.0{)} & 6.24 \\
 & J2R & 10.93 & -0.16 & 1.36 & 98.0 & {(}1.4{)} & 6.71 \\
\addlinespace[5pt]
\multicolumn{8}{@{}l}{\textbf{Generating assumption: J2R; true mean $\bm{11.38}$}} \\
\addlinespace[2pt]
Complete data, BB & --- & 11.06 & -0.32 & 1.25 & 90.0 & {(}3.0{)} & 4.72 \\
DK imputation, BB & MAR & 10.54 & -0.84 & 1.77 & 90.0 & {(}3.0{)} & 6.48 \\
 & Halfway & 10.70 & -0.68 & 1.55 & 93.0 & {(}2.6{)} & 6.24 \\
 & J2R & 10.93 & -0.45 & 1.42 & 98.0 & {(}1.4{)} & 6.71 \\
\bottomrule
\end{tabular}
\endgroup
\par\smallskip
\begin{minipage}{\linewidth}\footnotesize
Based on 100 replicates, with 50 imputations per rule. Means are standardized to the same 100 fixed baseline values. Analyses before introducing missingness use all generated scores. Retained outcomes are shared across generating assumptions, so a given imputation rule yields the same estimates but is evaluated against a different truth in each block. Coverage and mean width refer to nominal 95\% intervals: equal-tailed credible intervals for DK and confidence intervals for BB, with Rubin pooling after imputation. Coverage MCSEs are in percentage points.
\end{minipage}
\end{table}

\begin{table}[tbp]
\caption{Standardized mean difference at month~8 under matching generating and imputation assumptions ($n=100$).}
\label{tab:imputation-contrasts}
\centering\small
\begingroup
\setlength{\tabcolsep}{4pt}
\renewcommand{\arraystretch}{1.08}
\sisetup{group-digits = false, round-mode = none, table-align-text-pre = false, table-align-text-post = false}
\begin{tabular}{
@{}l
@{\hspace{14pt}}
S[table-format = -1.2]
@{\hspace{14pt}}
S[table-format = 1.2]
S[table-format = 1.2]
S[table-format = 2.1]
@{\hspace{\fontdimen2\font}}
S[table-format = 1.1, table-space-text-pre = {(}, table-space-text-post = {)}]
S[table-format = 1.2]
@{}}
\toprule
\textbf{Analysis} & {\textbf{Mean}} & {\textbf{Bias}} & {\textbf{RMSE}} & \multicolumn{2}{c}{\shortstack{\textbf{Coverage (\%)}\\\textbf{(MCSE)}}} & {\textbf{Width}} \\
\midrule
\multicolumn{7}{@{}l}{\textbf{MAR; true difference $\bm{-1.17}$}} \\
\addlinespace[2pt]
Complete data, BB & -1.03 & 0.14 & 1.62 & 96.0 & {(}2.0{)} & 6.45 \\
Observed data, DK & -0.99 & 0.18 & 2.11 & 99.0 & {(}1.0{)} & 9.10 \\
DK imputation, BB & -0.86 & 0.31 & 2.20 & 98.0 & {(}1.4{)} & 9.39 \\
\addlinespace[5pt]
\multicolumn{7}{@{}l}{\textbf{Halfway; true difference $\bm{-0.89}$}} \\
\addlinespace[2pt]
Complete data, BB & -0.79 & 0.11 & 1.62 & 96.0 & {(}2.0{)} & 6.49 \\
DK imputation, BB & -0.71 & 0.18 & 1.76 & 99.0 & {(}1.0{)} & 8.63 \\
\addlinespace[5pt]
\multicolumn{7}{@{}l}{\textbf{J2R; true difference $\bm{-0.61}$}} \\
\addlinespace[2pt]
Complete data, BB & -0.53 & 0.07 & 1.61 & 96.0 & {(}2.0{)} & 6.55 \\
DK imputation, BB & -0.49 & 0.11 & 1.33 & 99.0 & {(}1.0{)} & 8.32 \\
\bottomrule
\end{tabular}
\endgroup
\par\smallskip
\begin{minipage}{\linewidth}\footnotesize
The contrast is active minus reference, standardized to the same fixed baseline values. Based on 100 replicates, with 50 imputations under the generating rule. Coverage and mean width refer to nominal 95\% intervals: equal-tailed credible intervals for DK and confidence intervals for BB, with Rubin pooling after imputation. Coverage MCSEs are in percentage points.
\end{minipage}
\end{table}

\FloatBarrier
\section{Discussion}
\label{sec:discussion}

The proposed procedure extends the negative binomial Gaussian copula framework\cite{burger2026nb} to bounded scores and counts. DK margins preserve the finite integer support and provide explicit CDFs and quantiles for fitting and imputation. Reference borrowing is defined through a smooth median anchor, while subsequent analyses can target mean scores.

The interpretation of borrowing depends on both the margins and the retained history. With shapes shared across arms, J2R replaces the margin after the trigger with that in the reference arm at the same visit and covariate values. Conditional imputations remain informed by the active participant's retained outcomes. Halfway borrowing is defined on the logit scale of the normalized anchor and does not necessarily halve a treatment difference on the score scale.

The first simulation study supported parameter recovery under the generating model, with generally small bias and coverage close to nominal. Larger samples improved precision and reduced the modest intercept bias under greater concentration at zero. The imputation study considered one setting, in which borrowing shifted estimates in the expected direction and matching imputation gave empirical coverage at or above nominal. Modest downward bias in the active mean was also present in BB analyses before deletion, so it cannot be attributed solely to imputation. Under halfway borrowing and J2R, treatment differences had smaller absolute biases than active means, partly reflecting cancellation between arms.

The analysis model and variance estimator remain relevant to performance. Using the same marginal family for imputation and analysis does not by itself ensure compatibility, because averaging over dropout patterns can produce a distribution outside the specified DK regression family. Rubin's variance can also differ from the repeated sampling variance under reference borrowing.\cite{bartlett2023} The coverage findings therefore apply to the settings and analyses examined here.

In the BtheB application, estimated treatment differences favored BtheB numerically and moved toward zero at the final visit as borrowing increased. All confidence intervals for treatment differences included zero. The residual diagnostics identified local discrepancies, and no recorded score reached the upper bound, limiting the application's evidence about ceiling behavior.

The application illustrates sensitivity among participants with recorded follow-up, using a missing final tail as the trigger rather than a documented treatment event. As in other reference-based analyses, assumptions about unobserved outcomes cannot be verified from the recorded data and require clinical justification. The reported comparisons show how the treatment estimates respond to the specified borrowing assumptions within this analysis population.

\section*{Data and code availability}
The application data are available as \texttt{BtheB} in the R package \texttt{HSAUR2}.\cite{hsaur2}
\ifBlinded
The identifying code repository link is omitted for double anonymized peer review.
\else
R code for the application and simulation studies is available at \url{https://github.com/DABURGER1/RBMI-Discrete-Bounded}.

\section*{Conflict of interest}
\ConflictStatement

\section*{Ethics statement}
\EthicsStatement

\section*{Patient consent}
\ParticipationStatement

\section*{Funding}
\FundingStatement
\fi

\begingroup
\interlinepenalty=10000
\IfFileExists{SageV.bst}{\bibliographystyle{SageV}}{\bibliographystyle{sagev}}
\bibliography{bib}

@article{carpenter2013,
  author = {Carpenter, J. R. and Roger, J. H. and Kenward, M. G.},
  title = {{Analysis of longitudinal trials with protocol deviation: a framework for relevant, accessible assumptions, and inference via multiple imputation}},
  journal = {Journal of Biopharmaceutical Statistics},
  year = {2013},
  volume = {23},
  number = {6},
  pages = {1352--1371},
  doi = {10.1080/10543406.2013.834911}
}

@article{cro2020,
  author = {Cro, S. and Morris, T. P. and Kenward, M. G. and Carpenter, J. R.},
  title = {{Sensitivity analysis for clinical trials with missing continuous outcome data using controlled multiple imputation: a practical guide}},
  journal = {Statistics in Medicine},
  year = {2020},
  volume = {39},
  number = {21},
  pages = {2815--2842},
  doi = {10.1002/sim.8569}
}

@misc{burger2026nb,
  author = {Burger, D. A. and Lesaffre, E. and Martina, R.},
  title = {{Reference-based sensitivity analysis for repeated count outcomes using negative binomial margins and a Gaussian copula}},
  year = {2026},
  howpublished = {arXiv preprint arXiv:2609.31914},
  url = {https://arxiv.org/abs/2609.31914}
}

@article{barbiero2026dk,
  author = {Barbiero, A. and Hitaj, A.},
  title = {{A discrete Kumaraswamy distribution for modeling rating and ranking data}},
  journal = {Environmental and Ecological Statistics},
  year = {2026},
  volume = {33},
  number = {1},
  pages = {339--370},
  doi = {10.1007/s10651-025-00697-z}
}

@article{bayes2017,
  author = {Bayes, C. L. and Baz{\'a}n, J. L. and {de Castro}, M.},
  title = {{A quantile parametric mixed regression model for bounded response variables}},
  journal = {Statistics and Its Interface},
  year = {2017},
  volume = {10},
  number = {3},
  pages = {483--493},
  doi = {10.4310/SII.2017.v10.n3.a11}
}

@article{smith2012,
  author = {Smith, M. S. and Khaled, M. A.},
  title = {{Estimation of copula models with discrete margins via Bayesian data augmentation}},
  journal = {Journal of the American Statistical Association},
  year = {2012},
  volume = {107},
  number = {497},
  pages = {290--303},
  doi = {10.1080/01621459.2011.644501}
}

@article{brooks2017,
  author = {Brooks, M. E. and Kristensen, K. and {van Benthem}, K. J. and Magnusson, A. and Berg, C. W. and Nielsen, A. and Skaug, H. J. and M{\"a}chler, M. and Bolker, B. M.},
  title = {{glmmTMB balances speed and flexibility among packages for zero-inflated generalized linear mixed modeling}},
  journal = {The R Journal},
  year = {2017},
  volume = {9},
  number = {2},
  pages = {378--400},
  doi = {10.32614/RJ-2017-066}
}

@article{proudfoot2003,
  author = {Proudfoot, J. and Goldberg, D. and Mann, A. and Everitt, B. and Marks, I. and Gray, J. A.},
  title = {{Computerized, interactive, multimedia cognitive-behavioural program for anxiety and depression in general practice}},
  journal = {Psychological Medicine},
  year = {2003},
  volume = {33},
  number = {2},
  pages = {217--227},
  doi = {10.1017/S0033291702007225}
}

@manual{hsaur2,
  author = {Hothorn, T. and Everitt, B. S.},
  title = {{HSAUR2: A Handbook of Statistical Analyses Using R (2nd Edition)}},
  year = {2025},
  note = {R package version 1.1-21},
  url = {https://CRAN.R-project.org/package=HSAUR2}
}

@article{vehtari2021,
  author = {Vehtari, A. and Gelman, A. and Simpson, D. and Carpenter, B. and B{\"u}rkner, P.-C.},
  title = {{Rank-normalization, folding, and localization: an improved $\widehat{R}$ for assessing convergence of MCMC (with discussion)}},
  journal = {Bayesian Analysis},
  year = {2021},
  volume = {16},
  number = {2},
  pages = {667--718},
  doi = {10.1214/20-BA1221}
}

@article{dunn1996,
  author = {Dunn, P. K. and Smyth, G. K.},
  title = {{Randomized quantile residuals}},
  journal = {Journal of Computational and Graphical Statistics},
  year = {1996},
  volume = {5},
  number = {3},
  pages = {236--244},
  doi = {10.1080/10618600.1996.10474708}
}

@manual{dharma,
  author = {Hartig, F.},
  title = {{DHARMa: Residual Diagnostics for Hierarchical (Multi-Level / Mixed) Regression Models}},
  year = {2026},
  note = {R package version 0.5.0},
  url = {https://github.com/florianhartig/dharma}
}

@book{rubin1987,
  author = {Rubin, D. B.},
  title = {{Multiple Imputation for Nonresponse in Surveys}},
  publisher = {John Wiley \& Sons},
  address = {New York},
  year = {1987},
  doi = {10.1002/9780470316696}
}

@article{bartlett2023,
  author = {Bartlett, J. W.},
  title = {{Reference-based multiple imputation---what is the right variance and how to estimate it}},
  journal = {Statistics in Biopharmaceutical Research},
  year = {2023},
  volume = {15},
  number = {1},
  pages = {178--186},
  doi = {10.1080/19466315.2021.1983455}
}

@article{cro2025binary,
  author = {Cro, S. and Quartagno, M. and White, I. R. and Carpenter, J. R.},
  title = {{Reference-based multiple imputation for longitudinal binary data}},
  journal = {Statistics in Medicine},
  year = {2025},
  volume = {44},
  number = {3--4},
  pages = {e10301},
  doi = {10.1002/sim.10301}
}

@article{tang2018ordinal,
  author = {Tang, Y.},
  title = {{Controlled pattern imputation for sensitivity analysis of longitudinal binary and ordinal outcomes with nonignorable dropout}},
  journal = {Statistics in Medicine},
  year = {2018},
  volume = {37},
  number = {9},
  pages = {1467--1481},
  doi = {10.1002/sim.7583}
}

@article{keene2014,
  author = {Keene, O. N. and Roger, J. H. and Hartley, B. F. and Kenward, M. G.},
  title = {{Missing data sensitivity analysis for recurrent event data using controlled imputation}},
  journal = {Pharmaceutical Statistics},
  year = {2014},
  volume = {13},
  number = {4},
  pages = {258--264},
  doi = {10.1002/pst.1624}
}

@article{najera2019betabinomial,
  author = {Najera-Zuloaga, J. and Lee, D.-J. and Arostegui, I.},
  title = {{A beta-binomial mixed-effects model approach for analysing longitudinal discrete and bounded outcomes}},
  journal = {Biometrical Journal},
  year = {2019},
  volume = {61},
  number = {3},
  pages = {600--615},
  doi = {10.1002/bimj.201700251}
}

@article{cro2019anchored,
  author = {Cro, S. and Carpenter, J. R. and Kenward, M. G.},
  title = {{Information-anchored sensitivity analysis: theory and application}},
  journal = {Journal of the Royal Statistical Society: Series A (Statistics in Society)},
  year = {2019},
  volume = {182},
  number = {2},
  pages = {623--645},
  doi = {10.1111/rssa.12423}
}

@misc{padellini2018model,
  author = {Padellini, T. and Rue, H.},
  title = {{Model-aware quantile regression for discrete data}},
  year = {2018},
  howpublished = {arXiv preprint arXiv:1804.03714},
  url = {https://arxiv.org/abs/1804.03714}
}

@article{snowden2011,
  author = {Snowden, J. M. and Rose, S. and Mortimer, K. M.},
  title = {{Implementation of G-computation on a simulated data set: demonstration of a causal inference technique}},
  journal = {American Journal of Epidemiology},
  year = {2011},
  volume = {173},
  number = {7},
  pages = {731--738},
  doi = {10.1093/aje/kwq472}
}
\endgroup

\appendix
\titleformat{\section}
  {\normalfont\Large\bfseries}
  {Appendix \thesection}
  {1em}
  {}

\section{Quantiles and ordering}
\label{app:ordering}

Let $H_K$ denote the continuous extension of \eqref{eq:cdf}. It is strictly increasing on $\left(-1,K\right)$ and agrees with the discrete distribution function at every integer support point. For $0<u<1$, write $s_u=H_K^{-1}\left(u\right)$. The smallest integer $y$ with $F_K\left(y\right)\geq u$ is therefore $\left\lceil s_u\right\rceil$, giving \eqref{eq:inverse}. At $u=1/2$, substitution of \eqref{eq:parameterization} gives $s_{1/2}=\left(K+1\right)r-1$, proving \eqref{eq:median}.

For fixed $\phi$ and $u$, define
\begin{equation}
c_\phi\left(u\right)
=-\frac{1}{\phi}\log\left\{1-\left(1-u\right)^{1/b_\phi}\right\}>0,
\qquad
b_\phi=\frac{\log\left(1/2\right)}{\log\left\{1-\exp\left(-\phi\right)\right\}}.
\end{equation}
Since $a=-\phi/\log r$, the quantile becomes
\begin{equation}
Q\left(u;r,\phi\right)=
\left\lceil\left(K+1\right)r^{c_\phi\left(u\right)}-1\right\rceil.
\end{equation}
It is nondecreasing in $r$, and hence in $\eta=\logit\left(r\right)$. With a common latent probability $u$ and common shape across sensitivity rules, ordering their predictors orders their imputed scores. This also gives stochastic ordering of the margins. The result does not require an outcome-scale treatment contrast to be linear in the reference-borrowing weight $\omega$ defined in \mainsecref{sec:rules}, nor does it imply that a fitted analysis coefficient must follow the same ordering.

\label{LastMainPage}
\StartSupplementaryMaterial
\fancyfoot[C]{\small Page \thepage\ of \pageref*{LastPage}}

\section{Posterior parameter summaries}
\label{supp:parameters}

\begin{table}[H]
\caption{Posterior summaries of the DK Gaussian copula imputation model.}
\label{tab:dk-parameters}
\centering
\begingroup
\small
\setlength{\tabcolsep}{3pt}
\renewcommand{\arraystretch}{1.08}
\sisetup{table-align-text-pre = false, table-align-text-post = false,
  group-digits = false, output-decimal-marker = {.}, round-mode = none}
\ifdefined\DkCrIGap\else\newlength{\DkCrIGap}\fi
\ifdefined\DkCrIPadding\else\newlength{\DkCrIPadding}\fi
\ifdefined\DkCrICandidate\else\newlength{\DkCrICandidate}\fi
\setlength{\DkCrIPadding}{\maxdimen}
\setlength{\DkCrICandidate}{0pt}
\ifdim\DkCrICandidate<\DkCrIPadding
  \setlength{\DkCrIPadding}{\DkCrICandidate}\fi
\settowidth{\DkCrICandidate}{$-$}
\ifdim\DkCrICandidate<\DkCrIPadding
  \setlength{\DkCrIPadding}{\DkCrICandidate}\fi
\setlength{\DkCrIGap}{\dimexpr\fontdimen2\font - \DkCrIPadding\relax}
\begin{tabularx}{\linewidth}{@{}l>{\arraybackslash}X
S[table-format = -2.3]
S[table-format = -2.3, table-space-text-pre = {[}, table-space-text-post = {,}] @{\;}
S[table-format = -2.3, table-space-text-post = {]}]
S[table-format = -2.3]
S[table-format = -2.3, table-space-text-pre = {[}, table-space-text-post = {,}] @{\;}
S[table-format = -2.3, table-space-text-post = {]}]
@{}}
\toprule
\textbf{Parameter} & \textbf{Description} & \multicolumn{3}{c}{\textbf{TAU}} & \multicolumn{3}{c}{\textbf{BtheB}} \\
\cmidrule(lr){3-5}\cmidrule(l){6-8}
& & {\textbf{Mean}} & \multicolumn{2}{c}{\textbf{95\% CrI}} & {\textbf{Mean}} & \multicolumn{2}{c}{\textbf{95\% CrI}} \\
\midrule
\multicolumn{8}{@{}l}{\textbf{\textit{Arm-specific summaries at the centered baseline value}}} \\
\addlinespace[2pt]
$\alpha_{1} + \beta_{1} A_i$ & Intercept, month 2 & -0.947 & {[}-1.191{,} & -0.714{]} & -1.337 & {[}-1.601{,} & -1.086{]} \\
$\alpha_{2} + \beta_{2} A_i$ & Intercept, month 3 & -1.107 & {[}-1.399{,} & -0.823{]} & -1.549 & {[}-1.893{,} & -1.221{]} \\
$\alpha_{3} + \beta_{3} A_i$ & Intercept, month 5 & -1.485 & {[}-1.897{,} & -1.121{]} & -1.787 & {[}-2.207{,} & -1.388{]} \\
$\alpha_{4} + \beta_{4} A_i$ & Intercept, month 8 & -1.746 & {[}-2.229{,} & -1.334{]} & -1.921 & {[}-2.406{,} & -1.468{]} \\
\addlinespace[3pt]
$\xi_{i1}$ & Median anchor, month 2 & 16.922 & {[}13.915{,} & 20.040{]} & 12.366 & {[}9.744{,} & 15.148{]} \\
$\xi_{i2}$ & Median anchor, month 3 & 14.964 & {[}11.666{,} & 18.528{]} & 10.299 & {[}7.376{,} & 13.578{]} \\
$\xi_{i3}$ & Median anchor, month 5 & 10.938 & {[}7.350{,} & 14.737{]} & 8.302 & {[}5.344{,} & 11.783{]} \\
$\xi_{i4}$ & Median anchor, month 8 & 8.649 & {[}5.220{,} & 12.343{]} & 7.320 & {[}4.294{,} & 10.983{]} \\
\midrule
\multicolumn{8}{@{}l}{\textbf{\textit{Treatment contrasts and shared parameters}}} \\
\addlinespace[2pt]
\textbf{Parameter} & \textbf{Description} & \multicolumn{3}{c}{\textbf{Mean}} & \multicolumn{3}{c}{\textbf{95\% CrI}} \\
\cmidrule(l){1-8}
$\beta_{1}$ & Treatment coefficient, month 2 & \multicolumn{3}{c}{\tablenum[table-format = -1.3]{-0.389}} & \multicolumn{3}{c}{\mbox{\tablenum[table-format = -1.3, table-space-text-pre = {[}, table-space-text-post = {,}]{{[}-0.726{,}}\hspace{\DkCrIGap}\tablenum[table-format = -1.3, table-space-text-post = {]}]{-0.052{]}}}} \\
$\beta_{2}$ & Treatment coefficient, month 3 & \multicolumn{3}{c}{\tablenum[table-format = -1.3]{-0.442}} & \multicolumn{3}{c}{\mbox{\tablenum[table-format = -1.3, table-space-text-pre = {[}, table-space-text-post = {,}]{{[}-0.859{,}}\hspace{\DkCrIGap}\tablenum[table-format = -1.3, table-space-text-post = {]}]{-0.004{]}}}} \\
$\beta_{3}$ & Treatment coefficient, month 5 & \multicolumn{3}{c}{\tablenum[table-format = -1.3]{-0.302}} & \multicolumn{3}{c}{\mbox{\tablenum[table-format = -1.3, table-space-text-pre = {[}, table-space-text-post = {,}]{{[}-0.824{,}}\hspace{\DkCrIGap}\tablenum[table-format = -1.3, table-space-text-post = {]}]{0.255{]}}}} \\
$\beta_{4}$ & Treatment coefficient, month 8 & \multicolumn{3}{c}{\tablenum[table-format = -1.3]{-0.175}} & \multicolumn{3}{c}{\mbox{\tablenum[table-format = -1.3, table-space-text-pre = {[}, table-space-text-post = {,}]{{[}-0.797{,}}\hspace{\DkCrIGap}\tablenum[table-format = -1.3, table-space-text-post = {]}]{0.464{]}}}} \\
\addlinespace[3pt]
$\gamma_{11}$ & Baseline coefficient, month 2 & \multicolumn{3}{c}{\tablenum[table-format = -1.3]{0.447}} & \multicolumn{3}{c}{\mbox{\tablenum[table-format = -1.3, table-space-text-pre = {[}, table-space-text-post = {,}]{{[}0.283{,}}\hspace{\DkCrIGap}\tablenum[table-format = -1.3, table-space-text-post = {]}]{0.597{]}}}} \\
$\gamma_{21}$ & Baseline coefficient, month 3 & \multicolumn{3}{c}{\tablenum[table-format = -1.3]{0.605}} & \multicolumn{3}{c}{\mbox{\tablenum[table-format = -1.3, table-space-text-pre = {[}, table-space-text-post = {,}]{{[}0.393{,}}\hspace{\DkCrIGap}\tablenum[table-format = -1.3, table-space-text-post = {]}]{0.815{]}}}} \\
$\gamma_{31}$ & Baseline coefficient, month 5 & \multicolumn{3}{c}{\tablenum[table-format = -1.3]{0.646}} & \multicolumn{3}{c}{\mbox{\tablenum[table-format = -1.3, table-space-text-pre = {[}, table-space-text-post = {,}]{{[}0.383{,}}\hspace{\DkCrIGap}\tablenum[table-format = -1.3, table-space-text-post = {]}]{0.909{]}}}} \\
$\gamma_{41}$ & Baseline coefficient, month 8 & \multicolumn{3}{c}{\tablenum[table-format = -1.3]{0.533}} & \multicolumn{3}{c}{\mbox{\tablenum[table-format = -1.3, table-space-text-pre = {[}, table-space-text-post = {,}]{{[}0.227{,}}\hspace{\DkCrIGap}\tablenum[table-format = -1.3, table-space-text-post = {]}]{0.840{]}}}} \\
\addlinespace[3pt]
$\phi_{1}$ & Shape, month 2 & \multicolumn{3}{c}{\tablenum[table-format = -1.3]{2.349}} & \multicolumn{3}{c}{\mbox{\tablenum[table-format = -1.3, table-space-text-pre = {[}, table-space-text-post = {,}]{{[}1.930{,}}\hspace{\DkCrIGap}\tablenum[table-format = -1.3, table-space-text-post = {]}]{2.779{]}}}} \\
$\phi_{2}$ & Shape, month 3 & \multicolumn{3}{c}{\tablenum[table-format = -1.3]{2.022}} & \multicolumn{3}{c}{\mbox{\tablenum[table-format = -1.3, table-space-text-pre = {[}, table-space-text-post = {,}]{{[}1.639{,}}\hspace{\DkCrIGap}\tablenum[table-format = -1.3, table-space-text-post = {]}]{2.414{]}}}} \\
$\phi_{3}$ & Shape, month 5 & \multicolumn{3}{c}{\tablenum[table-format = -1.3]{2.055}} & \multicolumn{3}{c}{\mbox{\tablenum[table-format = -1.3, table-space-text-pre = {[}, table-space-text-post = {,}]{{[}1.604{,}}\hspace{\DkCrIGap}\tablenum[table-format = -1.3, table-space-text-post = {]}]{2.526{]}}}} \\
$\phi_{4}$ & Shape, month 8 & \multicolumn{3}{c}{\tablenum[table-format = -1.3]{2.161}} & \multicolumn{3}{c}{\mbox{\tablenum[table-format = -1.3, table-space-text-pre = {[}, table-space-text-post = {,}]{{[}1.637{,}}\hspace{\DkCrIGap}\tablenum[table-format = -1.3, table-space-text-post = {]}]{2.689{]}}}} \\
\addlinespace[3pt]
$\rho$ & One-month latent correlation & \multicolumn{3}{c}{\tablenum[table-format = -1.3]{0.774}} & \multicolumn{3}{c}{\mbox{\tablenum[table-format = -1.3, table-space-text-pre = {[}, table-space-text-post = {,}]{{[}0.703{,}}\hspace{\DkCrIGap}\tablenum[table-format = -1.3, table-space-text-post = {]}]{0.834{]}}}} \\
\bottomrule
\end{tabularx}
\endgroup
\par\smallskip
\begin{minipage}{\linewidth}
\footnotesize
Entries are posterior means and 95\% equal-tailed credible intervals (CrI), using all 40,000 post-warmup draws. Visit indices $j=1,\ldots,4$ correspond to months 2, 3, 5, 8. Arm-specific summaries use $B_i=0$, corresponding to a baseline BDI-II score of 23.1546. The intercept is $\alpha_j$ for TAU and $\alpha_j+\beta_j$ for BtheB. Median anchors $\xi_{ij}$ are in score units, before conditioning on retained follow-up outcomes; they are not the discrete medians or covariate-standardized means. Regression coefficients refer to the logit scale of the normalized median anchor. The baseline coefficient $\gamma_{j1}$ refers to a 10-point increase in baseline score. Baseline coefficients, shapes $\phi_j$, and $\rho$ are shared across arms. All imputation rules use this same fitted posterior.
\end{minipage}
\end{table}

\clearpage

\section{Residuals against fitted values}
\label{supp:residuals}

\suppfigref{fig:residuals-fitted} displays the DHARMa residuals against each recorded outcome's posterior predictive median, separately by visit and treatment. The residuals are the same as those summarized in \mainfigref{fig:residuals}. Locally weighted smooths describe patterns within treatment arms; they are descriptive curves without uncertainty bands. In particular, trends near the extremes of the fitted values may be influenced by few observations.

\begin{figure}[!htbp]
\centering
\includegraphics[width=\textwidth]{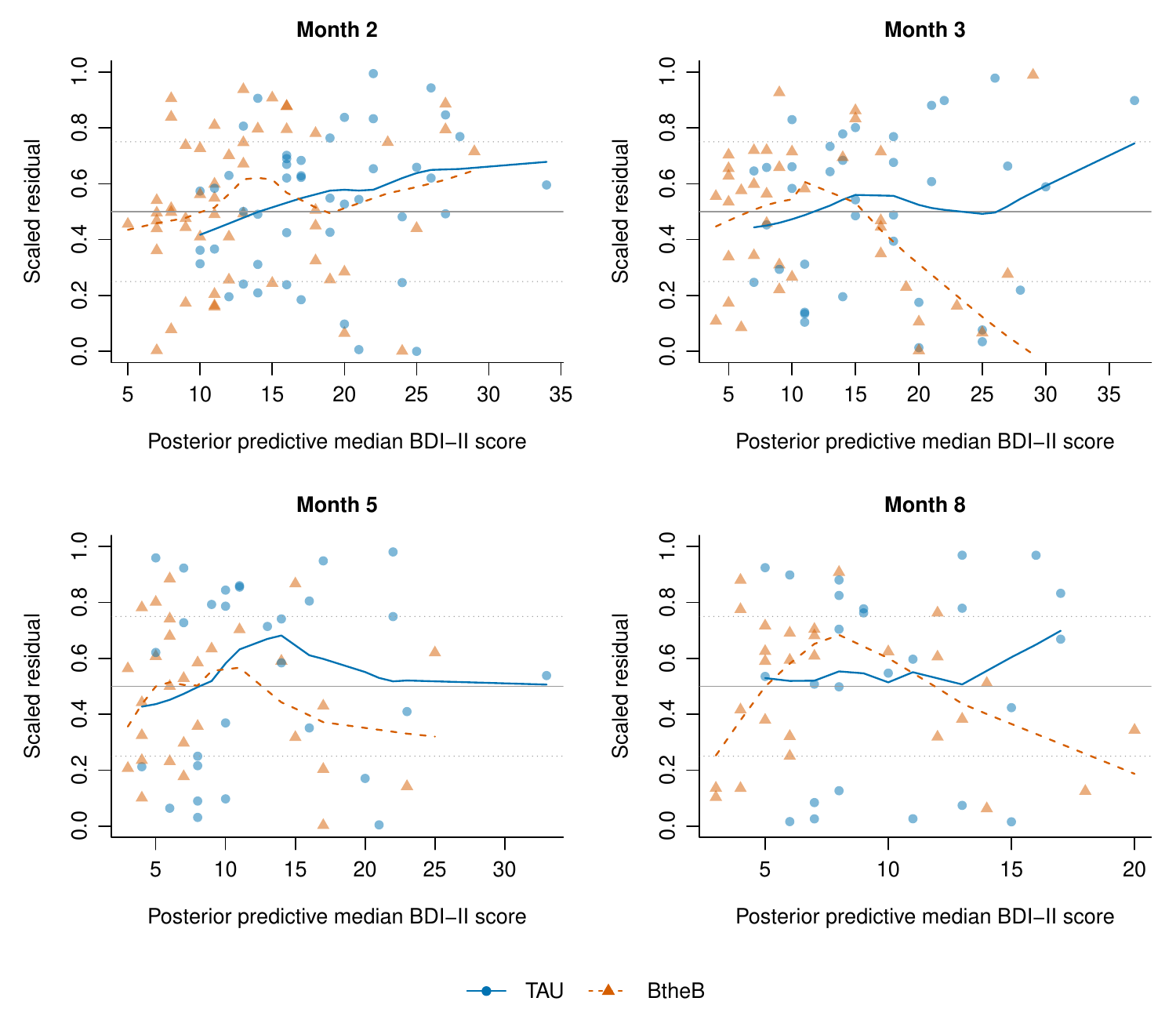}
\caption{DHARMa randomized predictive residuals against posterior predictive median BDI-II scores, by visit. Circles and solid curves denote TAU; triangles and dashed curves denote BtheB. Curves are locally weighted smooths. Horizontal lines mark 0.25, 0.50, and 0.75. All 40,000 posterior predictive replicates contribute to each residual.}
\label{fig:residuals-fitted}
\end{figure}

\clearpage
\stopcontents[supplement]

\fi
\end{document}